\documentclass[a4paper]{jpconf}
\usepackage{graphicx}
\usepackage{epsfig}
\def\H{H\hskip-8.5pt/\hskip2pt} 

\def\coeff#1#2{{\textstyle{#1\over #2}}}

\def\VEV#1{\left\langle #1\right\rangle}

\def\vev#1{\left\langle #1\right\rangle}

\def\lsim{\mathrel{\mathpalette\@versim<}}

\def\gsim{\mathrel{\mathpalette\@versim>}}

\def\Tr{{\rm Tr}\,}
\newcommand{\ba}{\begin{eqnarray}}
\newcommand{\ea}{\end{eqnarray}}
\newcommand{\lsp}{\tilde{\chi}}

\begin{document}
\title{Probing Models of Quantum Decoherence in Particle Physics and Cosmology
}

\author{Nikolaos E. Mavromatos and Sarben Sarkar}

\address{King's College London, Department of Physics, Theoretical Physics,  \\Strand London WC2R 2LS, U.K.

}

\ead{Nikolaos.Mavromatos@kcl.ac.uk, Sarben.Sarkar@kcl.ac.uk}

\begin{abstract}
In this review we first discuss the string theoretical motivations for
induced decoherence and deviations from ordinary quantum-mechanical
behaviour; this leads to intrinsic CPT violation in the context of an extended class of quantum-gravity models.
We then proceed to a description of
precision tests of CPT symmetry and quantum mechanics using mainly neutral kaons and neutrinos. We
also emphasize the possibly unique r\^ole of neutral meson factories in
providing specific tests of models where the quantum-mechanical
CPT operator is not well-defined, leading to modifications of
Einstein-Podolsky-Rosen particle correlators. Finally, we
discuss experimental probes of decoherence in a cosmological context,
including studies of dissipative relaxation models of dark energy in the context of non-critical (non-equilibrium) string theory and the associated modifications of the Boltzmann equation for the evolution of species abundances.

\end{abstract}


\section{Introduction: Decoherence, Quantum Gravity and CPT Violation}

A complete theory of {\it quantum gravity} (QG) will necessarily  address
fundamental issues, directly related to the emergence of space-time
and its structure at energies beyond the Planck energy scale $M_P
\sim 10^{19} $ GeV. From our experience with low-energy local
quantum field theories on flat space-times, we are tempted to expect
that a theory of QG should respect most of the
fundamental symmetries that govern the standard model of electroweak
and strong interactions, specifically Lorentz symmetry and CPT
invariance, i.e.  invariance under the combined action of Charge
Conjugation (C), Parity (P) and Time Reversal Symmetry (T).

CPT invariance is guaranteed in flat space-times by a theorem
applicable to any local relativistic quantum field theory of the type used
to describe currently the standard phenomenology of particle physics.
More precisely the \emph{ CPT theorem} states~\cite{cpt}:
\emph{Any quantum theory formulated on flat space-times is symmetric
under the combined action of CPT transformations, provided the
theory respects (i) Locality, (ii) Unitarity (i.e. conservation of
probability) and (iii) Lorentz invariance.}

The validity of any such theorem in the QG regime is an open and challenging issue since it is linked with our understanding of the nature of space-time at
(microscopic) Planckian distances $10^{-35}$~m. However there are reasons to believe that the CPT theorem \emph{may not} be valid (at least
in its strong form) in highly curved space-times with event horizons, such as those in the vicinity of 
black holes, or more generally in some QG models involving
{\it quantum space-time foam} backgrounds~\cite{wheeler}. The latter
are characterized by singular quantum fluctuations of space-time
geometry, such as microscopic black holes, {\it etc.}, with event horizons of
microscopic Planckian size. Such backgrounds result in {\it
apparent} violations of {\it unitarity} in the following sense:
there is some part of the initial information (quantum numbers of
incoming matter) which ``disappears'' inside the microscopic event
horizons, so that an observer at asymptotic infinity will have to
trace over such ``trapped'' degrees of freedom. One faces therefore
a situation where an initially pure state evolves in time and
becomes mixed. The asymptotic states are described by density
matrices, defined as
\begin{equation}
\rho _{\rm out} = {\rm Tr}_{M} |\psi ><\psi|~, \end{equation} where
the trace is over trapped (unobserved) quantum states that
disappeared inside the microscopic event horizons in the foam. Such
a non-unitary evolution makes it impossible to define a
standard quantum-mechanical scattering matrix. In ordinary local
quantum field theory, the latter connects asymptotic state vectors
in a scattering process
\begin{equation}
|{\rm out}> = S~|{\rm in}>,~S=e^{iH(t_f - t_i)}~,
\end{equation}
where $t_f - t_i$ is the duration of the scattering (assumed to be
much longer than other time scales in the problem, i.e.
${\rm lim}~t_i \to -\infty$, $t_f \to +\infty$). Instead, in
foamy situations, one can only define an operator that connects
asymptotic density matrices~\cite{hawking}:
\begin{equation}\label{dollar}
\rho_{\rm out} \equiv {\rm Tr}_{M}| {\rm out} ><{\rm out} | = \$
~\rho_{\rm in},~ \quad \$ \ne S~S^\dagger.
\end{equation}
The lack of factorization is attributed to the apparent loss of
unitarity of the effective low-energy theory, defined as the part of
the theory accessible to low-energy observers performing scattering
experiments. In such  situations particle phenomenology has to be
reformulated~\cite{ehns,poland} by viewing our low-energy world as
an open quantum system and using (\ref{dollar}). Correspondingly,
the usual Hamiltonian evolution of the wave function is replaced
by the Liouville equation for the density matrix~\cite{ehns}
\begin{equation}\label{evoleq}
\partial_t \rho = i[\rho, H] + \delta\H \rho~,
\end{equation}
where $\delta\H \rho $ is a correction of the form normally found
in open quantum-mechanical systems~\cite{lindblad}, although more general
forms are to be expected in QG~\cite{ms}.
This is what we denote by QG-induced {\it decoherence}, since the interaction with the quantum-gravitational environment  results in
quantum decoherence of the matter system, as is the case of open quantum mechanical systems in general~\cite{zurek,kiefer}.

The \$ matrix is {\it not invertible}, and this reflects the
effective unitarity loss. Since one of the requirements of CPT
theorem ( viz. unitarity) is violated there is no
CPT invariance (in theories which have CPT invariance in the absence of trapped states). This semi-classical analysis  leads to
more than a mere violation of the symmetry. The CPT
operator itself is \emph{not well-defined}, at least from an
effective field theory point of view. This is a strong form of CPT
violation (CPTV)~\cite{wald} and can be summarised by: \emph{In an open (effective) quantum
theory, interacting with an environment, e.g., quantum
gravitational, where} $ \$ \ne SS^\dagger $, \emph{CPT invariance is
violated, at least in its strong form}.
This form of violation introduces a fundamental arrow of
time/microscopic time irreversibility, unrelated in principle to CP
properties. Such
decoherence-induced CPT violation (CPTV) should occur in effective field
theories, since the low-energy experimenters do not have access to
all the degrees of freedom of QG (e.g., back-reaction
effects, \emph{etc.}). Some have conjectured that full CPT invariance
could be restored in the (still elusive) complete theory of QG.
In such a case, however, there may be~\cite{wald}
a \emph{weak form of CPT invariance}, in the sense of the possible existence
of \emph{decoherence-free subspaces} in the space of states of a
matter system. If this situation is realized, then the strong form
of CPTV will not show up in any measurable quantity (that
is, scattering amplitudes, probabilities \emph{etc.}).

The weak form of CPT invariance may be stated as follows: \emph{Let}
$\psi \in {\cal H}_{\it in}$, $\phi \in {\cal H}_{\it out}$
\emph{denote pure states in the respective Hilbert spaces ${\cal H}$
of in and out states, assumed accessible to experiment. If
$\theta $ denotes the (anti-unitary) CPT operator acting on pure
state vectors, then weak CPT invariance implies the following
equality between transition probabilities}
\begin{equation}
{\cal P}(\psi \to \phi) = {\cal P}(\theta^{-1}\phi \to \theta
\psi)~. \label{weakcpt}
\end{equation}
Experimentally, at least in principle,  it is possible to test
equations such as (\ref{weakcpt}), in the sense that, if decoherence
occurs, it induces (among other modifications) damping factors
in the time profiles of the corresponding transition probabilities.
The diverse experimental techniques for testing decoherence
range from
terrestrial laboratory experiments (in high-energy, atomic and
nuclear physics) to astrophysical observations of light from distant
extragalactic sources and high-energy cosmic neutrinos~\cite{poland}.

In the present article, we restrict ourselves to
decoherence and CPT invariance tests within the neutral kaon
system~\cite{ehns,lopez,huet,benatti,fide} and neutrinos~\cite{lisi,benattineutrino,icecube,barenboim,bm2,bmsw}. As we will argue later on,
this type of (decoherence-induced) CPTV exhibits
some fairly unique effects in $\phi$ factories~\cite{bmp},
associated with a possible modification of the
Einstein-Podolsky-Rosen (EPR) correlations of the entangled neutral
kaon states produced after the decay of the $\phi$-meson (similar
effects could be present for $B$ mesons produced in $\Upsilon$
decays).

We note for completeness two other possible mechanisms of CPTV in QG, which, however, shall not be discuss here, and which are {\it independent}
from decoherence. The first
is the {\it spontaneous breaking of Lorentz
symmetry (SBL)}~\cite{sme}; this type of CPTV does
not necessarily imply (nor does it invoke) decoherence.
In this case the ground state of the field theoretic
system is characterized by non-trivial vacuum expectation values of
certain tensorial quantities, $\langle {\cal A}_\mu \rangle \ne 0$~,
or $\langle {\cal B}_{\mu_1\mu_2\dots}\rangle \ne 0 $,
{\it etc.}
This may occur in (non-supersymmetric ground states of) string
theory and other models, such as loop QG~\cite{loops}.
The second mechanism for CPTV is associated with deviations from locality, e.g., as advocated in \cite{lykken}, in an attempt to explain observed
neutrino `anomalies', such as the LSND result~\cite{lsnd}, pointing towards suppressed flavour oscillations in the neutrino sector as compared to the antineutrino one.
Violations of
locality could also be tested with high precision, by studying
discrete symmetries in meson systems.

The reader should bear in mind that the
important difference between the CPTV in SBL models
and the CPTV due to the space-time foam is that
in the former case the CPT operator is well-defined, but \emph{does
not commute} with the effective Hamiltonian of the matter system. In
such cases one may parametrize the Lorentz and/or CPT breaking terms
by local field theory operators in the effective Lagrangian, leading
to a construction known as the ``standard model extension''
(SME)~\cite{sme}, which is a framework for studying precision tests of
such effects.

A note on the order of magnitude of such QG CPT-Violating effects
is in order.
If present, such effects are expected in general to be strongly suppressed, and thus difficult to detect experimentally, due to the weakness of gravity. Naively, QG has a dimensionful constant, $G_N \sim
1/M_P^2$, where $M_P =10^{19}$ GeV is the Planck scale. Hence, CPT
violating and decohering effects may be expected to be suppressed by
$E^3/M_P^2 $, where $E$ is a typical energy scale of the low-energy
probe. However, there could be cases where loop resummation and other
effects in theoretical models result in much larger CPT-violating
effects, of order $\frac{E^2}{M_P}$. This happens, for instance, in
some loop gravity approaches to QG~\cite{loops}, or some
non-equilibrium stringy models of space-time foam involving open
string excitations~\cite{emn}. Such large effects may lie within the
sensitivities of current or immediate future experimental facilities
(terrestrial and astrophysical), provided that enhancements due to
the near-degeneracy take place, as in the neutral-kaon case.
Moreover, possibly enhanced effects of QG decoherence may be encountered
in theories of space time with large extra dimensions, e.g. TeV scale
gravity~\cite{add}. In such cases, (high-energy) neutrinos appear the most sensitive probe to put stringent limits in various models in the foreseeable future~\cite{anchor}.

We next notice that, when interpreting experimental results in searches for CPT
violation, one should pay particular attention to disentangling
ordinary-matter-induced effects, that mimic CPTV, from
genuine effects due to QG~\cite{poland}. The
order of magnitude of matter induced effects, especially in neutrino
experiments, is often comparable to that expected in some models of
QG, and one has to exercise caution, by carefully
examining the dependence of the alleged ``effect'' on the probe
energy, or on the oscillation length (in neutrino oscillation
experiments). In most models, but \emph{not always}, since the
QG-induced CPTV is expressed as a back-reaction effect of
matter onto space-time, it increases with the probe energy  $E$ (and
oscillation length $L$ in the appropriate situations). In contrast,
ordinary matter-induced ``fake'' CPT-violating effects
decrease with $E$.

We emphasize that the phenomenology of CPTV is
complicated, and there does \emph{not} seem to be a \emph{single} figure
of merit for it. Depending on the precise way CPT might be violated
in a given model or class of models of QG, there are different ways
to test the violation~\cite{poland}.
In this review we describe
only some selected classes of such sensitive probes of CPT symmetry and
quantum-mechanical evolution (unitarity, decoherence): neutral mesons and
neutrinos.

The structure of the paper is as follows:
in the next section we discuss the basic ideas and the underlying formalism,
relevant for the  study of
the phenomenology of QG-induced decoherence.
In Section 3
we describe tests of
decoherence-induced CPTV using first (single-state) neutral kaon
systems, and then entangled neutral meson states. In this respect, we discuss the novel EPR-like modifications in
meson factories that may arise if the CPT operator is not well-defined,
as happens in some space-time foam models of QG.
We argue in favour of the unique character of such tests in
providing information on the stochastic nature of quantum
space-time, and we give some order-of-magnitude estimates within
some string-inspired  models. As we show, such models can be
falsified (or severely constrained) in the next-generation (i.e. upgraded) $\phi$-meson factories, such
as DA$\Phi$NE~\cite{dafne,adidomenico}. The enhancement of the effect provided by the
identical decay channels $(\pi^+\pi^-, \pi^+\pi^-)$ is unique.
In Section 4 we discuss
precision tests of decoherence effects in neutrino-oscillation experiments, which, with the exception of the above-mentioned EPR-correlation tests in meson factories, constitute the most sensitive particle-physics probes of QG-decoherence to date.
In section 5  we present a discussion on the r\^ole of decoherence in cosmology. In particular, we discuss dissipative cosmological models, which may arise in, say, string cosmology, whenever a non-equilibrium situation -due to a cosmically catastrophic event- appears.
In this context, we analyse the r\^ole of space-time boundaries (cosmic horizons) in possibly inducing decoherence of particle physics probes propagating in such space-time backgrounds. We examine dark-energy decoherent effects in flavour oscillations, within a specific non-critical string framework~\cite{emn,msdark}.
We finish our discussion with a brief description of non-equilibrium environmental effects on the Boltzmann equation for the evolution of the densities of cosmic relics, which may play a r\^ole in dark matter searches.
Conclusions and outlook at presented in section 6.

\section{Ideas and Methods for Quantum-Gravity Decoherence}
\label{sec:form}

\subsection{Non-critical string-framework for decoherence}

String theory is to date the most consistent theory of quantum gravity. 
In first-quantized string theory, de Sitter or other space-time backgrounds with cosmic horizons are {\it not conformal} on the world-sheet. The corresponding $\beta$-functions are non vanishing. Such non conformal backgrounds can be rendered conformal upon dressing the theory with the so-called Liouville mode~\cite{ddk}. The resulting $\sigma$-model world-sheet theory, then,
contains an extra target-space coordinate. In cases where the deviation from 
conformality is supercritical, the Liouville mode has a time-like signature, and can be identified with the target time~\cite{emn}. This formalism
provides a mathematically consistent way of incorporating de Sitter and 
other backgrounds with cosmological horizons in string theory. 

It can be shown, that the propagation of string matter in such non-conformal backgrounds is decoherent, the decoherence term being proportional to the world-sheet $\beta$-function. The following master equation for the evolution of stringy low-energy matter
in a non-conformal $\sigma $-model~can be derived\cite{emn}
\begin{equation}
\partial _{t}\rho =i\left[ \rho ,H\right] +:\beta ^{i}{\cal G}_{ij}\left[
g^{j},\rho \right] :  \label{master}
\end{equation}%
where $t$ denotes time (Liouville zero mode), the $H$ is the effective
low-energy matter Hamiltonian, $g^{i}$ are the quantum background target
space fields, $\beta ^{i}$ are the corresponding renormalization group $%
\beta $ functions for scaling under Liouville dressings and ${\cal G}_{ij}$
is the Zamolodchikov metric \cite{zam,kutasov} in the moduli space of the
string. To lowest order in the background field expansion the double colon symbol in (\ref{master}) represents the operator
ordering $:AB:=\left[ A,B\right] $ . The index $i$ labels the different
background fields as well as space-time. Hence the summation over $i,j$ in (%
\ref{master}) corresponds to a discrete summation as well as a covariant
integration $\int d^{D+1}y\,\sqrt{-g}$\bigskip\ where $y$ denotes a set of $%
\left( D+1\right) $-dimensional target space-time co-ordinates and $D$ is
the space-time dimensionality of the original non-critical string.

\begin{figure}[htb]
\centering
 \epsfig{file=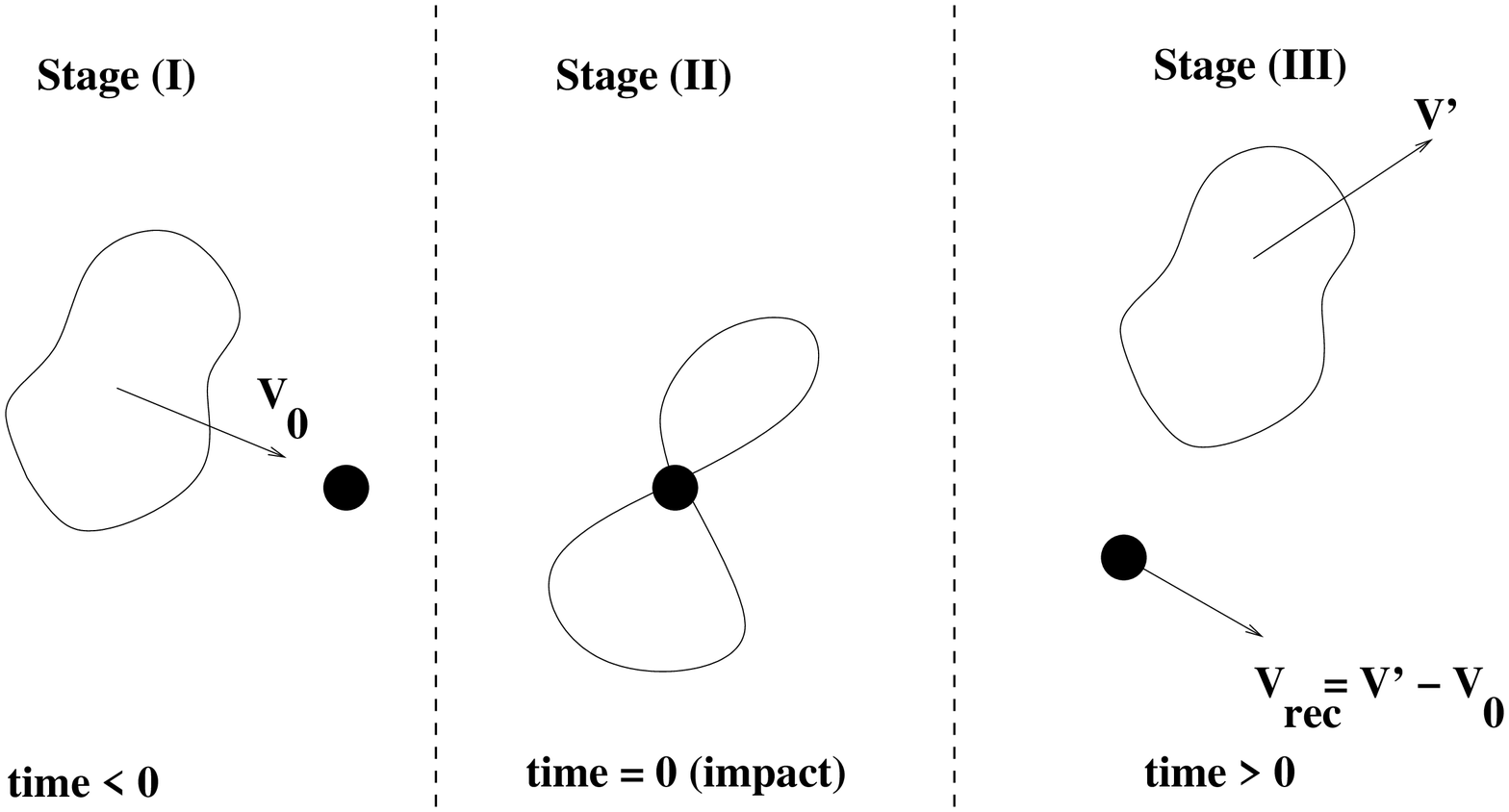, width=0.5\textwidth}
\hfill \epsfig{file=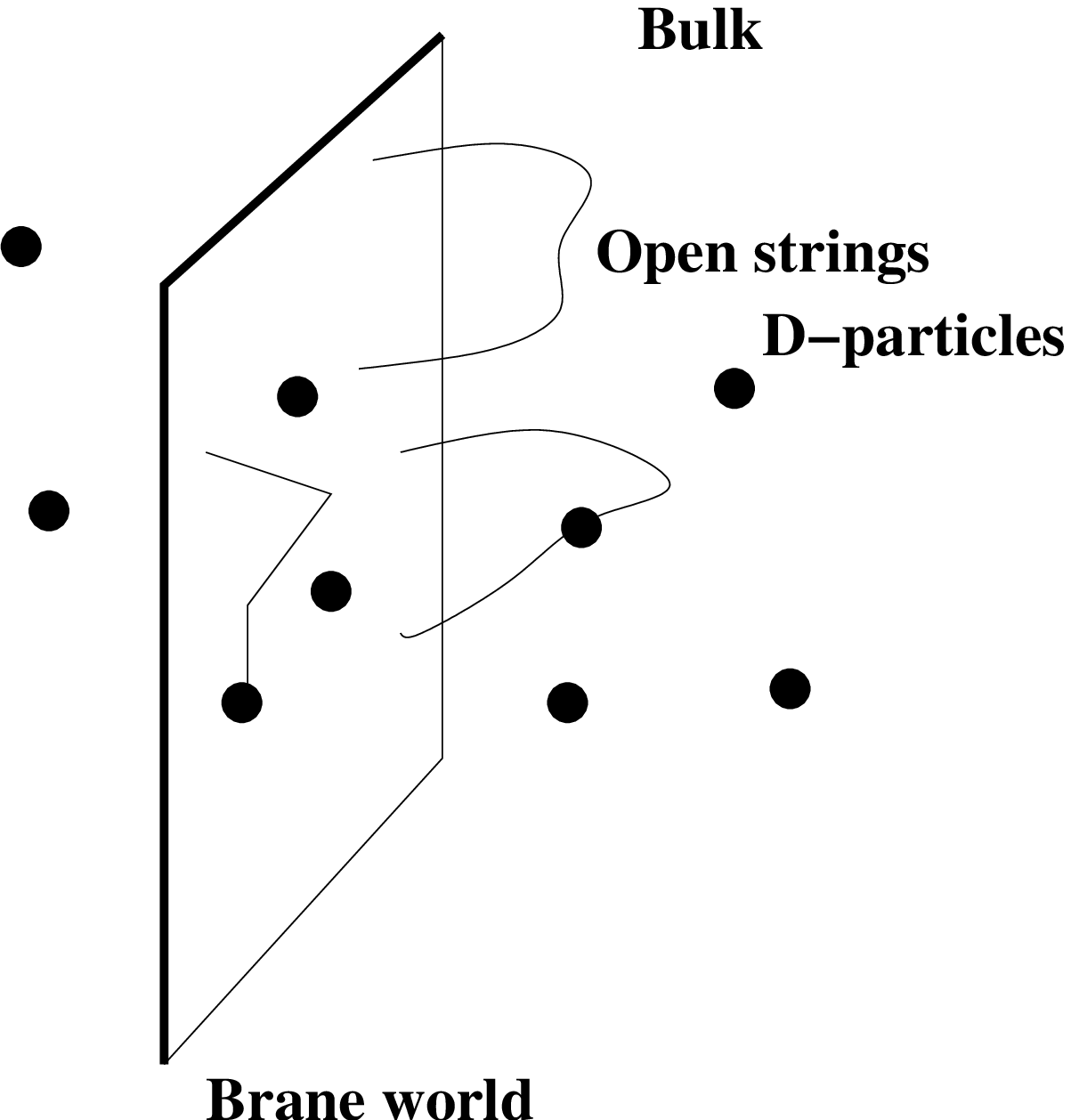, width=0.4\textwidth}
\caption{\underline{Left}: Recoil of closed string states with
D-particles (space-time defects). \underline{Right}: A
supersymmetric brane world model of D-particle foam. In both cases
the recoil of (massive) D-particle defect causes distortion of
space-time, stochastic metric fluctuations are possible and the
emergent post-recoil string state may differ by flavour and CP
phases.} \label{drecoil}\end{figure}

The discovery of new solitonic structures in superstring theory~\cite{polch}
has dramatically changed the understanding of target space structure. 
These
new non-perturbative objects are known as D-branes and their inclusion leads
to a scattering picture of space-time fluctuations.
As we have discussed previously, in this context one may consider
superstring models of space-time foam, containing a number of point-like 
solitonic structures (D-particles)~\cite{emw}. 
Heuristically, when low
energy matter given by a closed (or open) string propagating in a $\left(
D+1\right) $-dimensional space-time collides with a very massive D-particle
embedded in this space-time, the D-particle recoils as a result. Since there
are no rigid bodies in general relativity the recoil fluctuations of the
brane and their effectively stochastic back-reaction on space-time cannot be
neglected. On the brane there are closed and open strings propagating. Each
time these strings cross with a D-particle, there is a possibility of being
attached to it, as indicated in Fig. \ref{drecoil}. The entangled state
causes a back reaction onto the space-time, which can be calculated
perturbatively using logarithmic conformal field theory formalism~\cite{kmw}
. Now for large Minkowski time $t$, the non-trivial changes from the flat metric produced
from D-particle collisions are
\begin{equation}
g_{0i}\simeq \overline{u}_{i}\equiv \frac{u_{i}}{\varepsilon }\propto \frac{%
\Delta p_{i}}{M_{P}}  \label{recoil}
\end{equation}
where $u_i$ is the velocity and $\Delta p_{i}$ is the momentum transfer during a collision, $\varepsilon ^{-2}$ is identified with $t$ and $M_{P}$
is the Planck mass (actually, to be more precise $M_{P}=M_{s}/g_{s}$, where $%
g_{s}<1$ is the (weak) string coupling, and $M_{s}$ is a string mass scale);
so $g_{0i}$ is constant in space-time but depends on the energy content of
the low energy particle and the Ricci tensor $R_{MN}=0$ where $M$ and $N$
are target space-time indices. Since we are interested in fluctuations of
the metric the indices $i$ will correspond to the pair $M,N$. This master equation will serve as a framework for phenomenological applications.
 
\subsection{Stochastically Fluctuating Geometries, Light Cone
Fluctuations and Decoherence}

\begin{figure}[htb]
\centering
  \epsfig{file=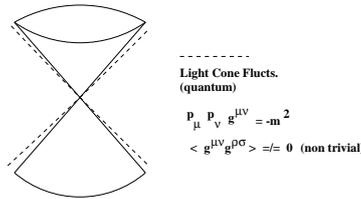, width=0.3\textwidth}
\caption{In stochastic space-time models of QG the
light cone may fluctuate, leading to decoherence and quantum
fluctuations of the speed of light in ``vacuo''.} \label{lcf}
\end{figure}

If the ground state of QG consists of ``fuzzy'' space-time, i.e.,
stochastically-fluctuating metrics, then a plethora of interesting
phenomena may occur, including light-cone
fluctuations~\cite{ford,emn} (c.f. Fig.~\ref{lcf}).
Such effects will lead to stochastic fluctuations in, say, arrival
times of photons with common energy, which can be detected with
high precision in astrophysical experiments~\cite{efmmn,ford}.
In addition, they may give rise to
decoherence of matter, in the sense of induced time-dependent
damping factors in the evolution equations of the (reduced) density
matrix of matter fields~\cite{emn,ms}.

Such ``fuzzy'' space-times are formally represented by metric
deviations which are fluctuating randomly about, say, flat Minkowski
space-time: $g_{\mu\nu} = \eta_{\mu\nu} + h_{\mu\nu}$, with $\langle
\cdots \rangle$ denoting statistical quantum averaging, and $\langle
g_{\mu\nu} \rangle = \eta_{\mu\nu} $  but $\langle h_{\mu\nu}(x)
h_{\lambda \sigma}(x') \rangle \ne 0 $,  i.e., one has only quantum
(light cone) fluctuations but not mean-field effects on dispersion
relations of matter probes. In such a situation Lorentz symmetry is
respected on the average, \emph{but} not in individual measurements.

The path of light follows null geodesics $ 0 = ds^2 =
g_{\mu\nu}dx^\mu dx^\nu $, with non-trivial fluctuations in geodesic
deviations, ${D^2n^{\mu}\over D\tau^2} =
-R^{\mu}_{\alpha\nu\beta}u^{\alpha}n^{\nu}u^{\beta}\,;$ in a
standard general-relativistic notation, $D/D\tau$ denotes the
appropriate covariant derivative operation,
$R^{\mu}_{\alpha\nu\beta}$ the (fluctuating) Riemann curvature
tensor, and $u^\mu$ ($n^\mu$) the tangential (normal) vector along
the geodesic.

Such an effect causes primarily fluctuations in the arrival time of
photons at the detector ($|\phi \rangle$=state of gravitons, $|0
\rangle$= vacuum state) {\small $$
 \Delta t_{obs}^2=|\Delta t_{\phi}^2-\Delta t_0^2 |=
{|\langle \phi| \sigma_1^2 |\phi\rangle-\langle 0| \sigma_1^2
|0\rangle|\over r^2}\equiv {|\langle \sigma_1^2 \rangle_R|\over r}\,
,
$$}
where {\small \begin{eqnarray}  && \langle \sigma_1^2 \rangle_R
=\frac{1}{8}(\Delta r)^2 \int_{r_0}^{r_1} dr \int_{r_0}^{r_1} dr'
\:\,
n^{\mu} n^{\nu} n^{\rho} n^{\sigma} \:\, \nonumber \\
&& \langle \phi| h_{\mu\nu}(x) h_{\rho\sigma}(x')+
 h_{\mu\nu}(x') h_{\rho\sigma}(x)|\phi \rangle \nonumber \end{eqnarray}}and
the two-point function of graviton fluctuations can be evaluated
using standard field theory techniques~\cite{ford}.

Apart from the stochastic metric fluctuations, however, the aforementioned
effects could also
induce decoherence of matter propagating in these types of
backgrounds~\cite{ms}, a possibility of particular interest for the
purposes of the present article.
Through the theorem of Wald~\cite{wald}, this implies that
the CPT operator is not well-defined, and hence one also has a
breaking of CPT symmetry.

We now proceed to describe briefly the general formalism used
for parametrizing such QG-induced decoherence, as far as the
CPT-violating effects on matter are concerned.

\subsection{Formalism for the Phenomenology of QG-induced Decoherence}

In this subsection we shall be very brief, giving
the reader a flavor of the formalism underlying such decoherent systems.
We shall discuss first a model-independent parametrization of
decoherence, applicable not only to QG media, but
covering a more general situation.

If the effects of the environment are such
that the modified evolution equation of the (reduced) density matrix
of matter $\rho$~\cite{kiefer} is linear, one can write
down a Lindblad evolution equation~\cite{lindblad}, provided that
(i) there is (complete) positivity of $\rho$, so that negative
probabilities do not arise at any stage of the evolution, (ii) the
energy of the matter system is conserved on the average, and (iii)
the entropy is increasing monotonically.

For $N$-level systems, the generic decohering Lindblad evolution for $\rho$ reads
\begin{eqnarray}\label{lindevol}
\frac{\partial \rho_\mu}{\partial t} = \sum_{ij} h_i\rho_j
{f}_{ij\mu} + \sum_{\nu} {L}_{\mu\nu}\rho_\nu~,\quad~\mu, \nu = 0, \dots N^2 -1, \quad i,j = 1, \dots N^2 -1~,
\end{eqnarray}
where the $h_i$ are Hamiltonian terms, expanded in an appropriate
basis, and the decoherence matrix $L$ has the form:
\begin{equation}
{L}_{0\mu}={L}_{\mu 0} =0~, \quad {L}_{ij} = \frac{1}{4}\sum_{k,\ell
,m} c_{l\ell}\left(-f_{i\ell m}f_{kmj} + f_{k i m}f_{\ell m
j}\right)~,
\label{cmatrix}
\end{equation}
 with $c_{ij}$ a {\it positive-definite matrix} and $f_{ijk}$
the structure constants of the appropriate $SU(N)$ group.
In this generic phenomenological description of decoherence, the
elements ${L}_{\mu\nu}$ are free parameters,
to be determined by experiment. We shall come back to this point in
the next subsection, where we discuss neutral kaon decays.

A rather characteristic feature of this equation is the
appearance of exponential damping, $e^{-(...)t}$, in interference terms
of the pertinent quantities (for instance, matrix elements $\rho$,
or asymmetries in the case of the kaon system, see below).
The exponents are proportional to (linear
combinations) of the elements of the decoherence
matrix~\cite{lindblad,ehns,kiefer}.

A specific example of Lindblad evolution is the propagation
of a probe in a medium with a stochastically fluctuating density~\cite{loreti}.
The formalism can be adapted to the case of stochastic space-time foam~\cite{ms,bmsw}.

The stochasticity of the space-time foam medium
is best described~\cite{bmsw} by
including in the time evolution of the neutrino
density matrix a a time-reversal (CPT) breaking
decoherence matrix of a double commutator form~\cite{loreti,bmsw},
\begin{eqnarray}
\partial_t \langle \rho\rangle =  L[\rho]~, \quad
L[\rho]=
-i[H + H'_{I},\langle \rho\rangle]-\Omega^2[H'_I,[H'_I,\langle \rho\rangle]]
\label{double}
\end{eqnarray}
where $\langle n(r) n(r') \rangle = \Omega^2n_0^2
\delta (r - r') $ denote the stochastic (Gaussian) fluctuations of
the density of the medium,
and
 \ba
H'_I=\left(%
\begin{array}{cc}
  (a_{\nu_{e}}-a_{\nu_{\mu}})\cos^2(\theta)  & (a_{\nu_{e}}-a_{\nu_{\mu}})\frac{\sin2\theta}{2}   \\
  (a_{\nu_{e}}-a_{\nu_{\mu}})\frac{\sin2\theta}{2}  & (a_{\nu_{e}}-a_{\nu_{\mu}})\sin^2(\theta)  \\
\end{array}%
\right) \ea
is the MSW-like interaction~\cite{msw}
in the mass eigenstate basis, where
$\theta$ is the mixing angle.
This double-commutator decoherence is a
specific case of Lindblad evolution \ref{lindevol}
which guarantees complete positivity of the time evolved
density matrix. For gravitationally-induced MSW effects (due to, say, black-hole foam
models as in \cite{bm2,ms}), one may denote the difference, between neutrino flavours,
of the effective interaction strengths, $a_i$,
 with the environment by:
\ba
\Delta a_{e\mu} \equiv a_{\nu_e}-a_{\nu_\mu} \propto G_N n_0
\ea
with $G_N=1/M_P^2$, $M_P \sim 10^{19}~{\rm GeV}$, the four-dimensional
Planck scale, and
in the case of
the gravitational MSW-like effect~\cite{bm2} $n_0$
represents the
density  of charge black hole/anti-black hole pairs.
This gravitational coupling replaces the weak interaction
Fermi coupling constant $G_F$ in the conventional MSW effect~\cite{msw}.

For two generation neutrino models, the corresponding oscillation probability $\nu_e \leftrightarrow
\nu_\mu$ obtained from (\ref{double}), in the small
parameter $\Omega^2 \ll 1$, which we assume here, as appropriate for the weakness of gravity fluctuations, reads to leading order:
{\small
\begin{eqnarray}
&&    P_{\nu_e\to \nu_{\mu}}=   \nonumber \\
    && \frac{1}{2} + e^{-\Delta a_{e\mu}^2\Omega^2t(1+\frac{\Delta_{12}^2}{4\Gamma}
(\cos(4\theta)-1))}
    \sin(t\sqrt{\Gamma})\sin^2(2\theta)\Delta
a_{e\mu}^2\Omega^2\Delta_{12}^2
    \left(\frac{3\sin^2(2\theta)\Delta_{12}^2}{4\Gamma^{5/2}}
-\frac{1}{\Gamma^{3/2}}\right) \nonumber \\
    && -e^{-\Delta
    a_{e\mu}^2\Omega^2t(1+\frac{\Delta_{12}^2}{4\Gamma}(\cos(4\theta)-1))}
    \cos(t\sqrt{\Gamma})
\sin^2(2\theta)\frac{\Delta_{12}^2}{2\Gamma}  \nonumber \\
    &&-e^{-\frac{\Delta a_{e\mu}^2\Omega^2 t \Delta_{12}^2\sin^2(2\theta)}{\Gamma}}
    \frac{(\Delta a_{e\mu}+\cos(2\theta)\Delta_{12})^2}{2\Gamma}
\label{2genprob}
\end{eqnarray}}
where $\Gamma= (\Delta a_{e\mu}\cos(2\theta)+\Delta_{12})^2+\Delta
a_{e\mu}^2\sin^2(2\theta)~,$ $\Delta_{12}=\frac{\Delta m_{12}^2}{2p}~.$

{}From (\ref{2genprob}) we easily conclude
that the exponents of the
damping factors due to the stochastic-medium-induced decoherence,
are of the generic form, for $t = L$, with $L$ the oscillation
length (in units of $c=1$):
\ba
{\rm exponent} \sim
-\Delta a_{e\mu}^2\Omega^2 t f(\theta)~;~
f(\theta) =
1+\frac{\Delta_{12}^2}{4\Gamma}(\cos(4\theta)-1)~, ~{\rm or} ~
\frac{\Delta_{12}^2\sin^2(2\theta)}{\Gamma}
\label{gammadelta}
\ea
that is proportional to the stochastic fluctuations of the
density of the medium.
The reader should note at this stage that, in
the limit $\Delta_{12}\to 0$, which could characterise the situation
in \cite{bm2}, where the space-time foam effects on the
induced neutrino mass difference are the dominant ones, the damping
factor is of the form $ {\rm exponent}_{{\rm gravitational~MSW}}
\sim -\Omega^2 (\Delta a_{e\mu})^2 L~,$ with the precise value of the
mixing angle $\theta$ not affecting the leading order of the various
exponents. However, in that case, as follows from (\ref{2genprob}),
the overall oscillation probability is suppressed by factors
proportional to $\Delta_{12}^2 $, and, hence, the stochastic
gravitational MSW effect~\cite{bm2}, although in principle
capable of inducing mass differences for neutrinos, however does not
suffice to produce the bulk of the oscillation probability, which is
thus attributed to conventional flavour physics.

We note now that the Lindblad type evolution is
\emph{not} the most generic evolution for QG models. In
cases of space-time foam corresponding to {\it stochastically
(random) fluctuating space-times}, such as the situations causing
light-cone fluctuations examined previously, there is a different
kind of decoherent evolution, with damping that is quadratic in
time, i.e., one has a $e^{-(...)t^2}$ suppression of interference
terms in the relevant observables. We now come to discuss this case.

A  specific  model  of  stochastic  space-time  foam  is  based  on  a
particular kind of gravitational foam~\cite{emn,emw,ms}, consisting of
``real''   (as   opposed  to   ``virtual'')   space-time  defects   in
higher-dimensional  space   times,  in  accordance   with  the  modern
viewpoint of our  world as a brane hyper-surface  embedded in the bulk
space-time~\cite{polch}.   This   model  is  quite   generic  in  some
respects, and we will use it  later to estimate the order of magnitude
of novel CPT violating effects in entangled states of kaons.

A model of space-time foam \cite{emw} can be based on a number
(determined by target-space supersymmetry) of parallel brane worlds
with three large spatial dimensions. These brane worlds
move in a bulk space-time,
containing a ``gas'' of point-like bulk branes, termed
``D-particles'', which are stringy space-time solitonic defects. One
of these branes is the observable Universe. For an observer on the
brane the crossing D-particles will appear as twinkling space-time
defects, i.e. microscopic space-time fluctuations. This will give
the four-dimensional brane world a ``D-foamy'' structure. Following
work on gravitational decoherence \cite{emn,ms}, the target-space
metric state, which is close to being flat, can be represented
schematically as a density matrix
\begin{equation}
\rho_{\mathrm{grav}}=\int d\,^{5}r\,\,f\left(  r_{\mu}\right)
\left| g\left(  r_{\mu}\right)  \right\rangle \left\langle g\left(
r_{\mu}\,\right)\right|  .\, \label{gravdensity}%
\end{equation}
\bigskip The parameters $r_{\mu}\,\left(  \mu=0,1 \ldots \right)  $ pertain
to appropriate space-time metric deformations and are
stochastic, with a Gaussian distribution $\,f\left(  r_{\mu}\,\right)
$
characterized by the averages%
\[
\left\langle r_{\mu}\right\rangle =0,\;\left\langle r_{\mu}r_{\nu
}\right\rangle =\Delta_{\mu}\delta_{\mu\nu}\,.
\]
We will assume that
the fluctuations of the metric felt by two entangled neutral mesons
are independent, and $\Delta_{\mu}\sim O\left(  \frac{E^{2}}%
{M_{P}^{2}}\right)  $, i.e., very small. As matter moves through the
space-time foam in a typical ergodic picture, the effect of time
averaging is assumed to be equivalent to an ensemble average.
For our present discussion we consider a
semi-classical picture for the metric, and therefore $\left|  g\left(
r_{\mu}\right)  \right\rangle $ in (\ref{gravdensity}) is a
coherent state.

In the specific model of foam discussed in \cite{ms}, there is a
recoil effect of the D-particle, as a result of its scattering with
stringy excitations that live on the brane world and represent
low-energy ordinary matter. As the space-time defects, propagating
in the bulk space-time, cross the brane hyper-surface from the bulk
in random directions, they scatter with matter. The associated
distortion of space-time caused by this scattering can be considered
dominant only along the direction of motion of the matter probe.
Random fluctuations are then considered about an average flat
Minkowski space-time. The result is an effectively  two-dimensional
approximate fluctuating metric describing the main effects~\cite{ms}

\begin{eqnarray}
g^{\mu\nu}= \left(\begin{array}{cc}
  -(a_1+1)^2 + a_2^2 & -a_3(a_1+1) +a_2(a_4+1) \\
  -a_3(a_1+1) +a_2(a_4+1) & -a_3^2+(a_4+1)^2 \\
\end{array}\right).
\label{flct}
 \end{eqnarray}
The $a_i$ represent the fluctuations and are assumed to be random
variables, satisfying $\langle a_i\rangle =0$ and  $\langle a_i a_j\rangle =
\delta_{ij}\sigma_i$,~$i,j=1,\dots 4$.

Such a (microscopic) model of space-time foam is not of Lindblad
type, as can be seen~\cite{ms} by considering the oscillation
probability for, say, two-level scalar systems describing
oscillating neutral kaons, $K^0 ~\leftrightarrow \overline{K}^0$.
In the approximation of small fluctuations one
finds the following form for the oscillation probability of the
two-level scalar system:
\begin{eqnarray}
\langle e^{i(\omega _{1}-\omega _{2})t}\rangle = \frac{4\tilde{d}^{2}}{(P_{1}P_{2})^{1/2}}\exp \left( \frac{\chi _{1}}{%
\chi _{2}}\right) \exp (i\tilde{b}t)\,, \label{timedep}
\end{eqnarray}%
where $\omega_i,~i=1,2$ are the appropriate energy levels~\cite{ms}
of the two-level kaon system in the background of the fluctuating
space-time~(\ref{flct}), and

{\small \begin{eqnarray*}
\chi _{1} &=&-4(\tilde{d}^{2}\sigma _{1}+\sigma _{4}k^{4})\tilde{b}%
^{2}t^{2}+2i\tilde{d}^{2}\widetilde{b}^{2}\widetilde{c}k^{2}\sigma
_{1}\sigma _{4}t^{3}, \\
\chi _{2} &=&4\tilde{d}^{2}-2i\tilde{d}^{2}(k^{2}\tilde{c}\sigma _{4}+2%
\tilde{b}\sigma _{1})t+ \nonumber \\
&& \widetilde{b}k^{2}\left( \widetilde{b}k^{2}-2%
\widetilde{d}^{2}\widetilde{c}\right) \sigma _{1}\sigma _{4}, \\
P_{1} &=&4\tilde{d}^{2}+2i\widetilde{d}\widetilde{b}\left( k^{2}-\widetilde{d%
}\right) \sigma _{2}t+\tilde{b}^{2}k^{4}\sigma _{2}\sigma _{3}t^{2}, \\
P_{2} &=&4\tilde{d}^{2}-2i\widetilde{d}^{2}\left(
k^{2}\widetilde{c}\sigma _{4}+2\widetilde{b}\sigma _{1}\right)
t+ \mathcal{O}\left( \sigma ^{2}\right)~,
\end{eqnarray*}}
with {\small \begin{eqnarray*}
\begin{array}{c}
\tilde{b}=\sqrt{k^{2}+m_{1}^{2}}-\sqrt{k^{2}+m_{2}^{2}}, \nonumber \\
\tilde{c}%
=m_{1}^{2}(k^{2}+m_{1}^{2})^{-3/2}-m_{2}^{2}(k^{2}+m_{2}^{2})^{-3/2}, \nonumber \\
\tilde{d}=\sqrt{k^{2}+m_{1}^{2}}\sqrt{k^{2}+m_{2}^{2}}.\nonumber
\end{array}
\end{eqnarray*}}
From this expression one can see~\cite{ms} that the stochastic
model of space-time foam leads to a modification of oscillation
behavior quite distinct from that of the Lindblad formulation. In
particular,
the transition probability displays a Gaussian time-dependence,
decaying as $e^{-(...)t^2}$, a modification of
the oscillation period, as well as additional power-law fall-off.

{}From this characteristic time-dependence, one can obtain bounds
for the fluctuation strength of space-time foam in particle-physics
systems, such as neutral mesons and neutrinos, which we restrict our attention to for the purposes of this presentation. When
discussing the CPTV effects of foam on entangled states and neutrinos we make use
of this specific model of stochastically fluctuating D-particle
foam~\cite{emw,ms}, in order to demonstrate the effects explicitly
and obtain definite order-of-magnitude estimates~\cite{bms,bmsw}.

\section{QG Decoherence and CPTV in Neutral Kaons}

\subsection{Single-Kaon Beam Experiments}

As mentioned in the previous subsection, QG may induce decoherence
and oscillations $K^0 \leftrightarrow {\overline
K}^0$~\cite{ehns,lopez}, thereby implying a two-level quantum
mechanical system interacting with a QG ``environment''. Adopting
the general assumptions of average energy conservation and monotonic
entropy increase, the simplest model for parametrizing decoherence
(in a rather model-independent way) is the (linear) Lindblad
approach mentioned earlier. Not all entries of a general
decoherence matrix are physical, and in order to isolate the
physically relevant entries one must invoke specific
assumptions, related to the symmetries of the particle system in
question. For the neutral kaon system, such an extra assumptions
is that the QG medium respects
the $\Delta S=\Delta Q$ rule.
In such a case, the modified Lindblad evolution equation
(\ref{evoleq}) for
the respective density matrices of neutral kaon matter can be
parametrized as follows~\cite{ehns}:
$$\partial_t \rho = i[\rho, H] + \delta\H \rho~,$$
where {\small $$H_{\alpha\beta}=\left( \begin{array}{cccc}  - \Gamma
& -\coeff{1}{2}\delta \Gamma
& -{\rm Im} \Gamma _{12} & -{\rm Re}\Gamma _{12} \\
 - \coeff{1}{2}\delta \Gamma
  & -\Gamma & - 2{\rm Re}M_{12}&  -2{\rm Im} M_{12} \\
 - {\rm Im} \Gamma_{12} &  2{\rm Re}M_{12} & -\Gamma & -\delta M    \\
 -{\rm Re}\Gamma _{12} & -2{\rm Im} M_{12} & \delta M   & -\Gamma
\end{array}\right) $$} and
$$ {\delta\H}_{\alpha\beta} =\left( \begin{array}{cccc}
 0  &  0 & 0 & 0 \\
 0  &  0 & 0 & 0 \\
 0  &  0 & -2\alpha  & -2\beta \\
 0  &  0 & -2\beta & -2\gamma \end{array}\right)~.$$
Positivity of $\rho$ requires: $\alpha, \gamma  > 0,\quad
\alpha\gamma>\beta^2$. Notice that $\alpha,\beta,\gamma$ violate
{\it both}  CPT, due to their decohering nature~\cite{wald}, and CP
symmetry, as they do not commute with the CP operator
$\widehat{CP}$~\cite{lopez}: $\widehat{CP} = \sigma_3 \cos\theta +
\sigma_2 \sin\theta$,$~~~~~[\delta\H_{\alpha\beta}, \widehat{CP} ]
\ne 0$.

An important remark is now in order. As pointed out in
\cite{benatti}, although the above parametrization is sufficient for
a single-kaon state to have a positive definite density matrix (and
hence probabilities) this is \emph{not} true when one considers
the evolution of entangled kaon states ($\phi$-factories).
In this latter case,
complete positivity is guaranteed only if
the further conditions
\begin{equation}\label{cons}
\alpha = \gamma~ {\rm and} ~\beta = 0
\end{equation}
are imposed. When incorporating entangled states, one should either
consider possible new effects (such as the $\omega$-effect
considered below) or apply the constraints (\ref{cons}) also to
single kaon states~\cite{benatti}. This is not necessarily the case
when other non-entangled particle states, such as neutrinos, are
considered, in which case the $\alpha,\beta,\gamma$ parametrization
of decoherence may be applied. Experimentally the complete
positivity hypothesis can be tested explicitly. In what follows, as
far as single-kaon states are concerned, we keep the
$\alpha,\beta,\gamma$ parametrization, and give the available
experimental bounds for them, but we always have in mind the
constraint (\ref{cons}) when referring to entangled kaon states in a
$\phi$-factory.

As already mentioned, when testing CPT symmetry with neutral kaons
one should be careful to distinguish two types of CPTV: {\bf (i)} CPTV
within Quantum Mechanics~\cite{fide}, leading to possible
differences between particle-antiparticle masses and widths: $\delta
m= m_{K^0} - m_{{\overline K}^0}$, $\delta \Gamma = \Gamma_{K^0}-
\Gamma_{{\overline K}^0} $. This type of CPTV could be,
for instance,
due to (spontaneous) Lorentz violation~\cite{sme}. In that
case the CPT operator is well-defined as a quantum mechanical
operator, but does not commute with the Hamiltonian of the system.
This, in turn, may lead to mass and width differences between particles
and antiparticles, among other effects. {\bf (ii)} CPTV through
decoherence~\cite{ehns,poland} via the parameters
$\alpha,\beta,\gamma$ (entanglement with the QG ``environment'',
leading to modified evolution for $\rho$ and $\$ \ne S~S^\dagger $).
In the latter case the CPT operator may not be well-defined, which
implies novel effects when one uses entangled states of kaons, as we
shall discuss in the next subsection.

\begin{figure}[htb]
\centering
  \epsfig{file=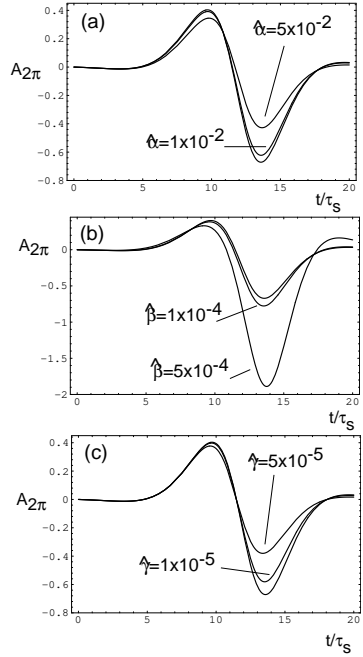, width=0.3\textwidth} \hfill
\caption{Neutral kaon decay asymmetries $A_{2\pi}$~\cite{lopez}, as a typical example indicating the effects of QG-induced decoherence.}
\label{AT}
\end{figure}

\begin{table}[thb]
\begin{center}
\begin{tabular}{lcc}
\underline{Process}&QMV&QM\\
$A_{2\pi}$&$\not=$&$\not=$\\
$A_{3\pi}$&$\not=$&$\not=$\\
$A_{\rm T}$&$\not=$&$=$\\
$A_{\rm CPT}$&$=$&$\not=$\\
$A_{\Delta m}$&$\not=$&$=$\\
$\zeta$&$\not=$&$=$
\end{tabular}
\caption{Qualitative comparison of predictions for various
observables in CPT-violating theories beyond (QMV) and within (QM)
quantum mechanics. Predictions either differ ($\not=$) or agree
($=$) with the results obtained in conventional quantum-mechanical
CP violation. Note that these frameworks can be qualitatively
distinguished via their predictions for $A_{\rm T}$, $A_{\rm CPT}$,
$A_{\Delta m}$, and $\zeta$.} \label{Table2}
\end{center}
\hrule
\end{table}

The important point to notice is that the two types of CPTV can be
{\it disentangled experimentally}~\cite{lopez}. The relevant
observables are defined as $ \VEV{O_i}= {\rm Tr}\,[O_i\rho] $. For
neutral kaons, one looks at decay asymmetries for $K^0, {\overline
K}^0$, defined as:
$$A (t) = \frac{
    R({\bar K}^0_{t=0} \rightarrow
{\bar f} ) -
    R(K^0_{t=0} \rightarrow
f ) } { R({\bar K}^0_{t=0} \rightarrow {\bar f} ) +
    R(K^0_{t=0} \rightarrow
f ) }~,$$ where $R(K^0\rightarrow f) \equiv \Tr[O_{f}\rho (t)]=$
denotes the decay rate into the final state $f$ (starting from a
pure $ K^0$ state at $t=0$).

In the case of neutral kaons, one may consider the following set of
asymmetries: (i) {\it identical final states}: $f={\bar f} = 2\pi $:
$A_{2\pi}~,~A_{3\pi}$, (ii) {\it semileptonic} : $A_T$ (final states
$f=\pi^+l^-\bar\nu\ \not=\ \bar f=\pi^-l^+\nu$), $A_{CPT}$
(${\overline f}=\pi^+l^-\bar\nu ,~ f=\pi^-l^+\nu$), $A_{\Delta m}$.
Typically, for instance when final states are $2\pi$, one has  a
time evolution of the decay rate $R_{2\pi}$: $ R_{2\pi}(t)=c_S\,
e^{-\Gamma_S t}+c_L\, e^{-\Gamma_L t} + 2c_I\, e^{-\Gamma
t}\cos(\Delta mt-\phi)$, where $S$=short-lived, $L$=long-lived,
$I$=interference term, $\Delta m = m_L - m_S$, $\Delta \Gamma =
\Gamma_S - \Gamma_L$, $\Gamma =\frac{1}{2}(\Gamma_S + \Gamma_L)$.
One may define the {\it decoherence parameter}
$\zeta=1-{c_I\over\sqrt{c_Sc_L}}$, as a (phenomenological) measure
of quantum decoherence induced in the system~\cite{fide}. For larger
sensitivities one can look at this parameter in the presence of a
regenerator~\cite{lopez}. In our decoherence scenario, $\zeta$
corresponds to a particular combination of the decoherence
parameters~\cite{lopez}:
$$ \zeta \to \frac{\widehat \gamma}{2|\epsilon ^2|} -
2\frac{{\widehat \beta}}{|\epsilon|}{\rm sin} \phi~,$$ with the
notation $\widehat{\gamma} =\gamma/\Delta \Gamma $, \emph{etc}.
Hence, ignoring the constraint (\ref{cons}), the best bounds on
$\beta$, or -turning the logic around- the most sensitive tests of
complete positivity in kaons, can be placed by implementing a
regenerator~\cite{lopez}.

The experimental tests (decay asymmetries) that can be performed in
order to disentangle decoherence from quantum-mechanical CPT
violating effects are summarized in Table \ref{Table2}. In Figure
\ref{AT} we give a typical profile of a
decay asymmetries~\cite{lopez}, from where bounds on QG decohering
parameters can be extracted. At present there are experimental bounds
available from CPLEAR measurements~\cite{cplear} $\alpha
< 4.0 \times 10^{-17} ~{\rm GeV}~, ~|\beta | < 2.3. \times 10^{-19}
~{\rm GeV}~, ~\gamma < 3.7 \times 10^{-21} ~{\rm GeV} $, which are
not much different from theoretically expected values in some
optimistic scenarios~\cite{lopez} $\alpha~,\beta~,\gamma = O(\xi
\frac{E^2}{M_{P}})$.

Recently, the experiment KLOE at DA$\Phi$NE updated these limits by
measuring for the first time the $\gamma$ decoherence parameter for
entangled kaon states~\cite{adidomenico}, as well as the (naive)
decoherence parameter $\zeta$ (to be specific, the KLOE
Collaboration has presented measurements for two $\zeta$ parameters,
one, $\zeta_{LS}$, pertaining to an expansion in terms of $K_L,K_S$
states, and the other, $\zeta_{0\bar 0}$, for an expansion in terms
of $K^0,\overline K^0$  states). We remind the reader once more
that, under the assumption of complete positivity for entangled
meson states~\cite{benatti}, theoretically there is only one
parameter to parametrize Lindblad decoherence, since $\alpha =
\gamma$, $\beta = 0$. In fact, the KLOE experiment has the greatest
sensitivity to this parameter $\gamma$. The latest KLOE
measurement~\cite{adidomenico} for $\gamma$ yields $\gamma_{\rm
KLOE} = (1.1^{+2.9}_{-2.4} \pm 0.4) \times 10^{-21}~{\rm GeV}$, i.e.
$\gamma < 6.4 \times 10^{-21}~{\rm GeV}$, competitive with the
corresponding CPLEAR bound~\cite{cplear} discussed above. It is
expected that this bound could be improved by an order of magnitude
in upgraded facilities, such as KLOE-2 at
DA$\Phi$NE-2~\cite{adidomenico}, where one expects $\gamma_{\rm
upgrade} \to \pm 0.2 \times 10^{-21} ~{\rm GeV}$.

The reader should also bear in mind that the Lindblad linear
decoherence is not the only possibility for a parametrization of QG
effects, see for instance the stochastically fluctuating space-time
metric approach discussed in Section 3.1 above. Thus, direct tests of
the complete positivity hypothesis in entangled states, and hence
the theoretical framework {\it per se}, should be performed by
independent measurements of all the three decoherence parameters
$\alpha,\beta,\gamma$; as far as we understand, such data
are currently available in kaon factories, but not yet analyzed in
detail~\cite{adidomenico}.

\subsection{CPTV and Modified EPR Correlations
of Entangled Neutral Kaon States}

\subsubsection{The $\omega$-Effect}

We now come to a description of an entirely novel effect~\cite{bmp}
of CPTV due to the ill-defined nature of the CPT
operator, which is \emph{exclusive} to neutral-meson factories, for
reasons explained below. The effects are qualitatively similar
for kaon and $B$-meson factories~\cite{bomega}, with the
important observation that in kaon factories there is a particularly
good channel, that of both correlated kaons decaying to $\pi^+\pi^-$.
In that channel the sensitivity of the effect increases because
the complex parameter $\omega$, parametrizing the relevant
EPR modifications~\cite{bmp}, appears in the particular
combination $|\omega|/|\eta_{+-}|$, with
$|\eta_{+-}| \sim 10^{-3}$. In the case of  $B$-meson factories
one should focus instead on the ``same-sign'' di-lepton
channel~\cite{bomega}, where high statistics is expected.

In this article we restrict ourselves to the case of
$\phi$-factories, referring the interested reader to the
literature~\cite{bomega} for the $B$-meson applications. We commence
our discussion by briefly reminding the reader of
EPR particle correlations.

The EPR effect was originally proposed as a {\it paradox}, testing the
foundations of Quantum Theory. There was the question whether
quantum correlations between spatially separated events implied
instant transport of information that would contradict special relativity.
It was eventually realized that no super-luminal propagation was
actually involved in the EPR phenomenon, and thus there was no
conflict with relativity.

\begin{figure}
\centering
  \epsfig{file=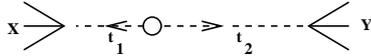, width=0.3\textwidth}
\caption{Schematic representation of the decay of a $\phi$-meson at
rest (for definiteness) into pairs of entangled neutral kaons, which
eventually decay on the two sides of the detector.}
\label{epr}\end{figure}

The EPR effect has been confirmed experimentally, e.g., in meson
factories: (i) a pair of particles can be created in a definite
quantum state, (ii) move apart and, (iii) eventually decay when they
are widely (spatially) separated (see Fig.~\ref{epr} for a schematic
representation of an EPR effect in a meson factory). Upon making a
measurement on one side of the detector and identifying the decay products,
we \emph{infer} the type of products appearing on the
other side; this is essentially the EPR correlation phenomenon.
It does \emph{not} involve any \emph{simultaneous measurement} on
both sides, and hence there is no contradiction with special
relativity. As emphasized by Lipkin~\cite{lipkin}, the EPR
correlations between different decay modes should be taken into
account when interpreting any experiment.

In the case of $\phi$ factories
it was \emph{claimed }~\cite{dunietz} that
due to EPR correlations,  \emph{irrespective} of CP, and CPT
violation, the \emph{final state} in $\phi$ decays: $e^+ e^-
\Rightarrow \phi \Rightarrow K_S K_L $ always contains $K_LK_S$
products.
This is a direct consequence
of imposing the requirement of {\it Bose statistics}
on the state $K^0 {\overline K}^0$ (to which the $\phi$ decays);
this, in turn, implies that the physical neutral meson-antimeson
state must be {\it symmetric} under C${\cal P}$, with C the charge
conjugation and ${\cal P}$ the operator that permutes the spatial
coordinates. Assuming {\it conservation} of angular momentum, and a
proper existence of the {\it antiparticle state} (denoted by a bar),
one observes that: for $K^0{\overline K}^0$ states which are
C-conjugates with C$=(-1)^\ell$ (with $\ell$ the angular momentum
quantum number), the system has to be an eigenstate of the
permutation operator ${\cal P}$ with eigenvalue $(-1)^\ell$. Thus,
for $\ell =1$: C$=-$ $\rightarrow {\cal P}=-$.  Bose statistics
ensures that for $\ell = 1$ the state of two \emph{identical} bosons
is \emph{forbidden}. Hence, the  initial entangled state:

{\scriptsize\begin{eqnarray*} &&|i> = \frac{1}{\sqrt{2}}\left(|K^0({\vec
k}),{\overline K}^0(-{\vec k})>
- |{\overline K}^0({\vec k}),{K}^0(-{\vec k})>\right)  \nonumber \\
&& = {\cal N} \left(|K_S({\vec k}),K_L(-{\vec k})> - |K_L({\vec
k}),K_S(-{\vec k})> \right)\nonumber
\end{eqnarray*}}with the normalization factor ${\cal N}=\frac{\sqrt{(1
+ |\epsilon_1|^2) (1 + |\epsilon_2|^2
)}}{\sqrt{2}(1-\epsilon_1\epsilon_2)} \simeq \frac{1 +
|\epsilon^2|}{\sqrt{2}(1 - \epsilon^2)}$, and
$K_S=\frac{1}{\sqrt{1 + |\epsilon_1^2|}}\left(|K_+> + \epsilon_1
|K_->\right)$, $K_L=\frac{1}{\sqrt{1 + |\epsilon_2^2|}}\left(|K_->
+ \epsilon_2 |K_+>\right)$, where $\epsilon_1, \epsilon_2$ are
complex parameters, such that $\epsilon \equiv \epsilon_1 +
\epsilon_2$ denotes the CP- \& T-violating parameter, whilst $\delta
\equiv \epsilon_1 - \epsilon_2$  parametrizes the CPT \& CP violation
within quantum mechanics~\cite{fide}, as discussed previously.
The $K^0 \leftrightarrow {\overline K}^0$ or $K_S
\leftrightarrow K_L$ correlations are apparent after evolution, at
any time $t > 0$ (with $t=0$ taken as the moment of the $\phi$ decay).

In the above considerations there is an implicit assumption,
which was noted in \cite{bmp}. The above arguments are valid
independently of CPTV, provided such violation occurs
within quantum mechanics, e.g., due to spontaneous Lorentz
violation, where the CPT operator is well defined.

If, however, CPT is \emph{intrinsically} violated, due, for instance,
to decoherence scenarios in space-time foam, then
the factorizability property of the
super-scattering matrix \$ breaks down, \$ $\ne SS^\dagger $, and
the generator of CPT is not well defined~\cite{wald}. Thus, the
concept of an ``antiparticle'' may be \emph{modified} perturbatively! The
perturbative modification of the properties of the antiparticle is
important, since the antiparticle state is a physical state which
exists, despite the ill-definition of the CPT operator. However, the
antiparticle Hilbert space will have components that are
\emph{independent} of the particle Hilbert
space.

In such a case,
the neutral mesons $K^0$ and ${\overline K}^0$ should \emph{no
longer} be treated as \emph{indistinguishable particles}. As a
consequence~\cite{bmp}, the initial entangled state in $\phi$
factories $|i>$, after the $\phi$-meson decay, will acquire a component
with opposite permutation (${\cal P}$) symmetry:

{\scriptsize \begin{eqnarray*} |i> &=& \frac{1}{\sqrt{2}}\left(|K_0({\vec
k}),{\overline K}_0(-{\vec k})>
- |{\overline K}_0({\vec k}),K_0(-{\vec k})> \right)\nonumber \\
&+&  \frac{\omega}{2} \left(|K_0({\vec k}), {\overline K}_0(-{\vec k})> + |{\overline K}_0({\vec
k}),K_0(-{\vec k})> \right)  \bigg]  \nonumber \\
& = & \bigg[ {\cal N} \left(|K_S({\vec
k}),K_L(-{\vec k})>
- |K_L({\vec k}),K_S(-{\vec k})> \right)\nonumber \\
&+&  \omega \left(|K_S({\vec k}), K_S(-{\vec k})> - |K_L({\vec
k}),K_L(-{\vec k})> \right)  \bigg]~, \nonumber
\end{eqnarray*}}where ${\cal N}$ is an appropriate normalization factor,
and $\omega = |\omega |e^{i\Omega}$ is a complex parameter,
parametrizing the intrinsic CPTV modifications of the EPR
correlations. Notice that, as a result of the $\omega$-terms, there
exist, in the two-kaon state,
$K_SK_S$ or $K_LK_L$ combinations,
which
entail important effects to the various decay channels. Due to this
effect, termed the $\omega$-effect by the authors of \cite{bmp},
there is \emph{contamination} of ${\cal P}$(odd) state with ${\cal P}$({\rm even})
terms. The $\omega$-parameter controls the amount of contamination
of the final ${\cal P}$(odd) state by the ``wrong'' (${\cal P}$(even)) symmetry state.

Later in this section  we will present a microscopic model where such a
situation is realized explicitly. Specifically,
an $\omega$-like effect appears due to the evolution in the space-time
foam, and the corresponding parameter turns out to be
purely imaginary and time-dependent~\cite{bms}.

\subsubsection{$\omega$-Effect Observables}

To construct the appropriate observable for the possible detection
of the $\omega$-effect, we consider the $\phi$-decay amplitude
depicted in Fig.~\ref{epr}, where one of the kaon products decays to
the  final state $X$ at $t_1$ and the other to the final state $Y$
at time $t_2$. We take $t=0$ as the moment of the $\phi$-meson
decay.

The relevant amplitudes read:
\begin{eqnarray*}
A(X,Y) = \langle X|K_S\rangle \langle Y|K_S \rangle \, {\cal N}
\,\left( A_1  +  A_2 \right)~, \nonumber
\end{eqnarray*}
with \begin{eqnarray*}
 A_1  &=& e^{-i(\lambda_L+\lambda_S)t/2}
[\eta_X  e^{-i \Delta\lambda \Delta t/2}
-\eta_Y  e^{i \Delta\lambda \Delta t/2}]\nonumber \\
A_2  &=&  \omega [ e^{-i \lambda_S t} - \eta_X \eta_Y e^{-i
\lambda_L t}] \nonumber
\end{eqnarray*}
denoting the CPT-allowed and CPT-violating parameters respectively,
and $\eta_X = \langle X|K_L\rangle/\langle X|K_S\rangle$ and $\eta_Y
=\langle Y|K_L\rangle/\langle Y|K_S\rangle$. In the above formulae, $t$
is the sum of the decay times $t_1, t_2$ and $\Delta t $ is their
difference (assumed positive).

\begin{figure}[htb]
\centering
  \epsfig{file=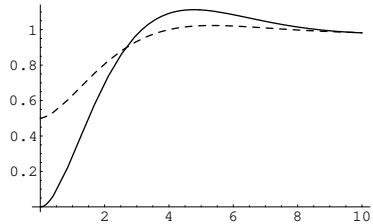, width=0.3\textwidth}
\caption{A characteristic case of the intensity $I(\Delta t)$, with
$|\omega|=0$ (solid line)  vs  $I(\Delta t)$ (dashed line) with
$|\omega|=|\eta_{+-}|$, $\Omega = \phi_{+-} - 0.16\pi$, for
definiteness~\cite{bmp}.} \label{intensomega}\end{figure}

The ``intensity'' $I(\Delta t)$ is the
desired \emph{observable} for a detection of the $\omega$-effect,
\begin{eqnarray*} I (\Delta t) \equiv \frac{1}{2} \int_{\Delta
t}^\infty dt\, |A(X,Y)|^2~. \nonumber
\end{eqnarray*}
depending only on $\Delta t$.

Its time profile reads~\cite{bmp}:

{\scriptsize \begin{eqnarray*} &&
I (\Delta t) \equiv \frac{1}{2} \int_{|\Delta t|}^\infty dt\,
|A(\pi^+\pi^-,\pi^+\pi^-)|^2  = \nonumber
\\ && |\langle\pi^{+}\pi^{-}|K_S\rangle|^4 |{\cal N}|^2 |\eta_{+-}|^2
\bigg[ I_1  + I_2  +  I_{12} \bigg]~,\nonumber
\end{eqnarray*}}where

{\scriptsize \begin{eqnarray*} && I_1 (\Delta t) =
\frac{e^{-\Gamma_S \Delta t} + e^{-\Gamma_L \Delta t} - 2
e^{-(\Gamma_S+\Gamma_L) \Delta t/2} \cos(\Delta m \Delta t)}
{\Gamma_L+\Gamma_S}
\nonumber \\
&& I_2 (\Delta t) =  \frac{|\omega|^2 }{|\eta_{+-}|^2}
\frac{e^{-\Gamma_S \Delta t} }{2 \Gamma_S}
\nonumber \\
&& I_{12} (\Delta t) = - \frac{4}{4 (\Delta m)^2 + (3 \Gamma_S +
\Gamma_L)^2}  \frac{|\omega|}{|\eta_{+-}|} \times
\nonumber \\
&&\bigg[ 2 \Delta m \bigg( e^{-\Gamma_S \Delta t} \sin(\phi_{+-}-
\Omega) - \nonumber \\
&&  e^{-(\Gamma_S+\Gamma_L) \Delta t/2} \sin(\phi_{+-}- \Omega
+\Delta m \Delta t)\bigg)
\nonumber \\
&&  - (3 \Gamma_S + \Gamma_L) \bigg(e^{-\Gamma_S \Delta t}
\cos(\phi_{+-}- \Omega) - \nonumber \\
&& e^{-(\Gamma_S+\Gamma_L) \Delta t/2} \cos(\phi_{+-}- \Omega
+\Delta m \Delta t)\bigg)\bigg]~, \nonumber
\end{eqnarray*}}with $\Delta m = m_S - m_L$ and $\eta_{+-}= |\eta_{+-}|
e^{i\phi_{+-}}$ in the usual notation~\cite{fide}.

A typical case for the relevant intensities, indicating clearly the
novel CPTV $\omega$-effects, is depicted in Fig.~\ref{intensomega}.

As announced, the novel $\omega$-effect appears in
the combination $\frac{|\omega|}{|\eta_{+-}|}$, thereby implying
that the decay channel to $\pi^+\pi^-$ is particularly sensitive to
the $\omega$ effect~\cite{bmp}, due to the enhancement by
$1/|\eta_{+-}| \sim 10^{3}$, implying sensitivities up to
$|\omega|\sim 10^{-6}$ in $\phi$ factories. The physical reason for
this enhancement is that $\omega$ enters through $K_SK_S$ as opposed to
$K_LK_S$ terms, and the $K_L \to \pi^+\pi^-$ decay is CP-violating.

\subsubsection{Microscopic Models for the $\omega$-Effect and Order-of-Magnitude
Estimates}

For future experimental searches for the $\omega$-effect it is
important to estimate its expected order of magnitude, at least  in
some models of foam.

A specific model is that of the D-particle foam~\cite{emw,ms,bms},
discussed already in connection with the stochastic
metric-fluctuation approach to decoherence. An important feature for
the appearance of an $\omega$-like effect is that, during each
scattering with a D-particle defect, there is (momentary) capture of
the string state (representing matter) by the defect, and a possible
change in phase and/or flavour for the particle state emerging from
such a capture (see Fig.~\ref{drecoil}).

 The induced metric distortions,
including such flavour changes for the emergent post-recoil matter
state, are:
\begin{eqnarray}
&& g^{00} =\left( -1+r_{4}\right) \mathsf{1}~, \quad
g^{01}=g^{10}=r_{0}\mathsf{1}+ r_{1}\sigma_{1}+ r_{2}\sigma_{2}
+r_{3}\sigma_{3}, \quad
 g^{11} =\left( 1+r_{5}\right) \mathsf{1} \end{eqnarray} where
the $\sigma_i$ are Pauli matrices.

The target-space metric state is the density matrix $\rho_{\mathrm{grav}}$
defined at (\ref{gravdensity})~\cite{bms}, with the same assumptions for
the parameters $r_{\mu}$ stated there.
The order of magnitude  of the metric elements $g_{0i}\simeq
\overline{v}_{i,rec} \propto g_s\frac{\Delta p_{i}}{M_{s}} $, where
$\Delta p_{i}\sim {\tilde \xi} p_i $ is the momentum transfer during
the scattering of the particle probe (kaon) with the D-particle
defect, $g_{s}<1$ is the string coupling, assumed weak, and $M_{s}$
is the string scale, which in the modern approach to string/brane
theory is not necessarily identified with the four-dimensional
Planck scale, and is left as a phenomenological parameter to be
constrained by experiment.

To estimate the order of magnitude of the $\omega$-effect we
construct the gravitationally-dressed initial entangled state using
stationary perturbation theory for degenerate states~\cite{bmp}, the
degeneracy being provided by the CP-violating effects. As
Hamiltonian function we use

{\scriptsize \begin{eqnarray*}
\widehat{H}=g^{01}\left( g^{00}\right)
^{-1}\widehat{k}-\left( g^{00}\right) ^{-1}\sqrt{\left(
g^{01}\right) ^{2}{k}^{2}-g^{00}\left( g^{11}k^{2}+m^{2}\right)  }
\nonumber
\end{eqnarray*}}describing propagation in the above-described
stochastically-fluctuating space-time. To leading order in the
variables $r$ the interaction Hamiltonian reads:
\begin{equation}
\widehat{H_{I}} = -\left(  { r_{1} \sigma_{1} + r_{2} \sigma_{2}}
\right) \widehat{k} \nonumber
\end{equation}
with the notation {\small $\left| K_{L}\right\rangle =\left|
\uparrow\right\rangle~, \quad \left| K_{S}\right\rangle =\left|
\downarrow\right\rangle .$} The gravitationally-dressed initial states
then can be constructed using stationary perturbation theory:

{\scriptsize
\begin{eqnarray*} \left|  k^{\left(  i\right)
},\downarrow\right\rangle _{QG}^{\left( i\right)  } =  \left| k^{\left(
i\right) },\downarrow\right\rangle ^{\left( i\right)  } + \left| k^{\left(
i\right) },\uparrow\right\rangle ^{\left( i\right) } \alpha^{\left(
i\right)  }~, \nonumber
\end{eqnarray*}}where {\small $ \alpha^{\left(  i\right)  }= \frac{^{\left( i\right)
}\left\langle \uparrow, k^{\left( i\right) }\right| \widehat{H_{I}}\left|
k^{\left(  i\right)  }, \downarrow\right\rangle ^{\left(  i\right)
}}{E_{2} - E_{1}} $}. For $\left|  { k^{\left(  i\right) }, \uparrow}
\right\rangle ^{\left( i \right)  } $  the dressed state is obtained by
$\left| \downarrow\right\rangle \leftrightarrow\left|
\uparrow\right\rangle $ and $\alpha\to\beta$ where  {\small $
\beta^{\left( i\right) }= \frac{^{\left( i\right)  }\left\langle
\downarrow, k^{\left( i\right)  }\right| \widehat{H_{I}}\left| k^{\left(
i\right)  }, \uparrow\right\rangle ^{\left(  i\right) }}{E_{1} - E_{2}}$}.

The totally antisymmetric ``gravitationally-dressed'' state of two mesons
(kaons) is then:

{\scriptsize
\begin{eqnarray*}
\begin{array}
[c]{l}%
\left|  {k, \uparrow} \right\rangle _{QG}^{\left(  1 \right)  }
\left|  { - k, \downarrow} \right\rangle _{QG}^{\left(  2 \right)  }
- \left|  {k, \downarrow} \right\rangle _{QG}^{\left(  1 \right)  }
\left|  { - k, \uparrow} \right\rangle _{QG}^{\left(  2 \right)  }
= \\
\left|  {k, \uparrow} \right\rangle ^{\left(  1 \right) } \left|
{ - k, \downarrow} \right\rangle ^{\left(  2 \right)  } - \left| {k,
\downarrow} \right\rangle ^{\left(  1 \right)  } \left| { - k,
\uparrow} \right\rangle
^{\left(  2 \right)  }\\
+  \left|  {k, \downarrow} \right\rangle ^{\left(  1 \right) }
\left|  { - k, \downarrow} \right\rangle ^{\left(  2 \right)  }
\left(  {\beta^{\left(  1 \right)  } - \beta^{\left(  2 \right)  } }
\right) + \\
\left|  {k, \uparrow} \right\rangle ^{\left( 1 \right)
} \left|  { - k, \uparrow} \right\rangle ^{\left(  2 \right)  }
\left( {\alpha^{\left(  2 \right)  } - \alpha^{\left(
1 \right)  } } \right) \\
 + \beta^{\left(  1 \right)  } \alpha^{\left(  2 \right) }
\left|  {k, \downarrow} \right\rangle ^{\left(  1 \right) } \left| {
- k, \uparrow} \right\rangle ^{\left(  2 \right)  } - \alpha^{\left(
1 \right)  } \beta^{\left( 2 \right)  } \left|  {k, \uparrow}
\right\rangle ^{\left(  1
\right)  } \left|  { - k, \downarrow} \right\rangle ^{\left(  2 \right)  }~.\\
\label{entangl}%
\end{array}\end{eqnarray*}}Notice here
that, for our order-of-magnitude estimates, it suffices to assume that the
initial entangled state of kaons is a pure state. In practice, due to the
omnipresence of foam, this may not be entirely true, but this should not
affect our order-of-magnitude estimates based on such an assumption.

With these remarks in mind we then write for the initial state of
two kaons after the $\phi$ decay:
{\scriptsize  \begin{eqnarray*}
&&\left| \psi\right\rangle = \left|  k,\uparrow\right\rangle
^{\left( 1\right) }\left| -k,\downarrow\right\rangle ^{\left(
2\right) }-\left| k,\downarrow\right\rangle ^{\left(  1\right)
}\left|
-k,\uparrow \right\rangle ^{\left(  2\right)  }+
 \xi \left| k,\uparrow\right\rangle ^{\left( 1\right) }\left|
-k,\uparrow\right\rangle ^{\left(  2\right) }+  \xi^{\prime} \left|
k,\downarrow\right\rangle ^{\left( 1\right) }\left|
-k,\downarrow\right\rangle ^{\left( 2\right)  }~,
\end{eqnarray*}}where for $r_{i} \propto\delta_{i1} $ we have
$\xi= \xi^{\prime}$, that is strangeness violation, whilst for
$r_{i} \propto\delta_{i2}$ $\longrightarrow $ $\xi= -\xi^{\prime}$)
(since $\alpha^{\left( i \right) } = \beta^{\left( i \right) } )$ we
obtain a strangeness conserving $\omega$-effect.

Upon averaging the density matrix over $r_{i}$, only the
$|\omega|^{2}$ terms survive: {\small \begin{eqnarray*}
|\omega|^{2} = \mathcal{O}\left(  \frac{1}{(E_{1} - E_{2})^2}
(\langle \downarrow, k |H_{I} |k, \uparrow\rangle)^{2} \right)  \sim
\frac{\Delta_{2} k^{2}}{(m_{1} - m_{2})^{2}} \nonumber
\end{eqnarray*}}for momenta of order of the rest energies, as is the case of a
$\phi$ factory.

Recalling that in the recoil D-particle model under consideration we
have~\cite{emn,bms} $\Delta_{2} = {\tilde \xi}^{2} k^{2}/M_{P}^{2}$,
we obtain the following order of magnitude estimate of the $\omega$
effect: {\small \begin{eqnarray} |\omega|^{2} \sim\frac{{\tilde
\xi}^{2} k^{4}}{M_{P}^{2} (m_{1} - m_{2})^{2}}~. \label{orderomega}
\end{eqnarray}}For neutral kaons with momenta of the order of the rest energies
$|\omega| \sim10^{-4} |{\tilde \xi}|$. For $1 >
{\tilde\xi}\ge10^{-2}$ this not far below the sensitivity of current
facilities, such as KLOE at DA$\Phi$NE. In fact, the KLOE experiment
has just released the first measurement of the $\omega$
parameter~\cite{adidomenico}: {\small \begin{eqnarray*} &&
{\rm Re}(\omega) = \left(
1.1^{+8.7}_{-5.3} \pm 0.9\right)\times 10^{-4}~, \qquad
{\rm Im}(\omega) = \left( 3.4^{+4.8}_{-5.0} \pm 0.6\right)\times
10^{-4}~. \nonumber\end{eqnarray*}}One can constrain the $\omega$
parameter (or, in the context of the above specific model, the
momentum-transfer parameter ${\tilde \xi}$) significantly in
upgraded facilities. For instance, there are the following
perspectives for KLOE-2 at (the upgraded)
DA$\Phi$NE-2~\cite{adidomenico}: ${\rm Re}(\omega),~ {\rm
Im}(\omega) \longrightarrow 2 \times 10^{-5}$.

Let us now mention that $\omega$-like effects can also be generated by
the Hamiltonian evolution of the system as a result of gravitational
medium interactions. To this end, let us consider the Hamiltonian
evolution in our stochastically-fluctuating D-particle-recoil
distorted space-times, {\scriptsize  $ \left| \psi\left(  t\right)
\right\rangle =\exp\left[ -i\left( \widehat
{H}^{(1)}+\widehat{H}^{\left( 2\right) }\right)
\frac{t}{\hbar}\right] \left| \psi\right\rangle$}.

Assuming for simplicity $\xi = \xi^\prime = 0$, it is easy to
see~\cite{bms} that the time-evolved state of two kaons contains
strangeness-conserving $\omega$-terms:
\begin{eqnarray*}  \left|  \psi\left(  t\right)  \right\rangle \sim
e^{-i\left(  \lambda_{0}^{\left(  1\right)  }+
\lambda_{0}^{\left(  2\right)  }\right)  t}%
 \varpi\left(  t\right) \times
\left\{  \left| k,\uparrow\right\rangle ^{\left( 1\right) }\left|
-k,\uparrow\right\rangle ^{\left( 2\right)  }-\left|
k,\downarrow\right\rangle ^{\left(  1\right) }\left|  -k,\downarrow
\right\rangle ^{\left(  2\right)  }\right\}~. \nonumber
\end{eqnarray*}
The quantity $\varpi (t)$ obtained within this specific model is purely imaginary,
\begin{eqnarray*} {\cal O}\left(\varpi\right)  \simeq
i\frac{2\Delta_{1}^{\frac{1}{2}}k}{\left(
k^{2}+m_{1}^{2}\right)^{\frac{1}{2}}-\left(
k^{2}+m_{2}^{2}\right)^{\frac{1}{2}}}\times \cos\left( \left| \lambda^{\left(  1\right) }\right| t\right)
\sin\left( \left| \lambda^{\left(  1\right) }\right| t\right) =
\varpi_{0}\sin\left( 2\left|  \lambda^{\left( 1\right) }\right|
t\right), \nonumber
\end{eqnarray*} with $\Delta_{1}^{1/2}\sim\left| {\tilde \xi}\right| \frac{\left| k\right|
}{M_{P}}$,
$\varpi_{0}\equiv\frac{\Delta_{1}^{\frac{1}{2}}%
k}{\left(  k^{2}+m_{1}^{2}\right)  ^{\frac{1}{2}}-\left(  k^{2}+m_{2}%
^{2}\right)  ^{\frac{1}{2}}}$, $\left|  \lambda^{\left(  1\right)
}\right|  \sim\left( 1+\Delta_{4}^{\frac{1}{2}}\right)
\sqrt{\chi_{1}^{2}+\chi_{3}^{2}}$,~ $\chi_{3}\sim\left(
k^{2}+m_{1}^{2}\right)  ^{\frac{1}{2}}-\left(
k^{2}+m_{2}^{2}\right) ^{\frac{1}{2}}$.

It is important to notice the time dependence of the medium-generated
effect. It is also interesting to observe that, if in the initial state
we have  a strangeness-conserving (-violating) combination,
$\xi = -\xi^\prime $  ($\xi = \xi^\prime $),
then the time
evolution generates time-dependent strangeness-violating
(-conserving $\omega$-) imaginary effects.

The above description of medium effects using Hamiltonian evolution is
approximate, but suffices for the purposes of obtaining order-of-magnitude
estimates for the relevant parameters. In the complete description of the
above model  there is of course decoherence~\cite{bms,emn}, which affects
the evolution and induces mixed states for kaons. A complete analysis of
both effects, $\omega$-like and decoherence in entangled neutral kaons of
a $\phi$-factory, has already been carried out~\cite{bmp}, with the upshot
that the various effects can be disentangled experimentally, at least in
principle.

Finally, as the analysis of
\cite{bms} demonstrates, no $\omega$-like effects are generated by
thermal bath-like (rotationally-invariant, isotropic) space-time
foam situations, argued to simulate the QG
environment in some models \cite{garay}. In this
way, the potential observation of an $\omega$-like effect in
EPR-correlated meson states would in principle distinguish various
types of space-time foam.

\subsubsection{Disentangling the $\omega$-Effect from
the C(even)  Background and Decoherent
Evolution Effects}

When interpretating experimental results on delicate violations of CPT
symmetry, it is important
to disentangle (possible) genuine effects from those
due to ordinary physics.
Such a situation arises in connection with the $\omega$-effect,
that must be disentangled from the C(even) background
characterizing the decay products in a
$\phi$-factory~\cite{dunietz}.  The C(even) background $e^+e^- \Rightarrow 2\gamma \Rightarrow K^0
{\overline K}^0$ leads to states of the form
\begin{eqnarray*}
|b> = |K^0 {\overline K}^0 (C({\rm even}))> =
\frac{1}{\sqrt{2}}\left(K^0({\vec k}) {\overline K}^0 (-{\vec k})
+  {\overline K}^0({\vec k}) K^0 (-{\vec k}) \right)~, \nonumber
\end{eqnarray*} which at first sight mimic the $\omega$-effect, as such states would
also produce contamination by terms $K_SK_S,~K_LK_L$.

Closer inspection reveals, however,
that the two types of effects can be clearly disentangled
experimentally~\cite{bmp,poland} on two accounts:
(i) the expected magnitude of the two effects,
in view of the the above-described estimates in QG models and the unitarity bounds that suppress the `fake' (C(even)-background) effect~\cite{dafne2},
and (ii) the
different way the genuine QG-induced $\omega$-effect interferes with the
the C(odd) background~\cite{bmp}.

Finally, we remark that it is also possible to disentangle~\cite{bmp}
the $\omega$-effect from decoherent evolution
effects~\cite{bmp}, due to their different structure.
For instance, the experimental disentanglement of $\omega$ from the decoherence
parameter $\gamma$ in completely-positive QG-Lindblad models of entangled neutral mesons~\cite{ehns,lopez,benatti} is possible as a result of different symmetry properties and different structures generated by the time evolution
of the pertinent terms.

\section{Neutrino Tests of QG Decoherence }

Neutrino oscillations is one of the most sensitive particle-physics probes of QG-decoherence to date. The presence of QG decoherence effects would affect the profiles of the oscillation probabilities among the various neutrino flavours,
not only by the presence of appropriate {\it damping factors} in front of the
interference oscillatory terms, but also by appropriate modifications of the oscillation period and phase shifts of the oscillation arguments~\cite{lisi,benattineutrino,icecube,poland}.

In the context of our present talk, we shall restrict ourselves to
a brief discussion of a three-generation fit, including QG-decoherence effects, to all presently available data, including LSND results~\cite{lsnd}.
For more details we refer the reader to the literature~\cite{bmsw}.

We assume for simplicity a Lindblad environment, for which the evolution
(\ref{lindevol} applies. However, as we have discussed in section \ref{sec:form}, a
combination of both Lindblad environment and stochastically fluctuating
space-time backgrounds (\ref{flct}),
affecting also the Hamiltonian parts of the matter-density-matrix evolution (\ref{lindevol}), could also be in place.
Without a detailed microscopic model, in the three generation case, the precise physical
significance of the decoherence matrix cannot be fully understood.
In \cite{bmsw} we considered the simplified case in which
the matrix $C \equiv (c_{kl})$ (\ref{cmatrix}) is {\it assumed} to be
of the form
 \begin{eqnarray}
  C =   \left(%
\begin{array}{cccccccc}
  c_{11} & 0 & 0 & 0 & 0 & 0 & 0 & 0 \\
  0 & c_{22} & 0 & 0 & 0 & 0 & 0 & 0 \\
  0 & 0 & c_{33} & 0 & 0 & 0 & 0 & c_{38} \\
  0 & 0 & 0 & c_{44} & 0 & 0 & 0 & 0 \\
  0 & 0 & 0 & 0 & c_{55} & 0 & 0 & 0 \\
  0 & 0 & 0 & 0 & 0 & c_{66} & 0 & 0 \\
  0 & 0 & 0 & 0 & 0 & 0 & c_{77} & 0 \\
  0 & 0 & c_{38} & 0 & 0 & 0 & 0 & c_{88} \\
\end{array}%
\right)
\label{cmatrix2}
 \end{eqnarray}
Positivity of the evolution can be guaranteed if and only if
the matrix $C$ is positive
and hence has non-negative eigenvalues.
As discussed in detail in \cite{ms,bmsw}, such special choices can be realised for models of
the propagation of neutrinos in  models of stochastically
fluctuating environments~\cite{loreti}, where the decoherence term
corresponds to an appropriate double commutator involving operators that entangle with the environment.The quantum-gravity space time
foam may in principle behave as one such stochastic
environment~\cite{poland,ms,lopez}, and it is this point of view
that we critically examined in \cite{bmsw} and review here, in the context of the entirety of the presently available neutrino data.

The simplified form of the $c_{ij}$ matrix given in (\ref{cmatrix2})
implies a matrix  $L_{ij}$ in (\ref{lindevol}) of the form:
 \begin{eqnarray}
    L = \left(%
\begin{array}{cccccccc}
  {\cal D}_{11} & -\Delta_{12} & 0 & 0 & 0 & 0 & 0 & 0 \\
  \Delta_{12} & {\cal D}_{22} & 0 & 0 & 0 & 0 & 0 & 0 \\
  0 & 0 & {\cal D}_{33} & 0 & 0 & 0 & 0 & {\cal D}_{38} \\
  0 & 0 & 0 & {\cal D}_{44} & -\Delta_{13} & 0 & 0 & 0 \\
  0 & 0 & 0 & \Delta_{13} & {\cal D}_{55} & 0 & 0 & 0 \\
  0 & 0 & 0 & 0 & 0 & {\cal D}_{66} & -\Delta_{23} & 0 \\
  0 & 0 & 0 & 0 & 0 & \Delta_{23} & {\cal D}_{77} & 0 \\
  0 & 0 & {\cal D}_{83} & 0 & 0 & 0 & 0 & {\cal D}_{88} \\
\end{array}%
\right)
 \end{eqnarray}
where we have used the notation
$\Delta_{ij}=\frac{m_i^2-m_j^2}{2p}$, and the matrix ${\cal D}_{ij}$
is expressed in terms of the elements of the $C$-matrix:
${\cal D}_{ij}=\sum_{k,l,m} \frac{c_{kl}}{4}\left(-f_{ilm}f_{kmj}
    +f_{kim}f_{lmj}\right)~.$

The probability of a
neutrino of flavor $\nu_{\alpha}$, created at time $t=0$, being converted to a
flavor $\nu_{\beta}$ at a later time t, is calculated in the
Lindblad framework~\cite{lindblad} to be
 \ba
    P_{\nu_{\alpha}\rightarrow \nu_{\beta}}(t)=\Tr [\rho^{\alpha}(t)\rho^{\beta}]
    = \frac{1}{3}+ \frac{1}{2}\sum_{i,j,k}
    e^{\lambda_{k}t}D_{ik}D_{kj}^{-1}
    \rho_{j}^{\alpha}(0)\rho_{i}^{\beta}~.
\label{lindbladprob}
 \ea
where $\lambda_k $ denote the eigenvalues of the Lindblad-decoherence matrix $L_{ij}$.

A detailed analysis~\cite{bmsw} gives the the final expression for the probability as:
 \ba
    \nonumber P_{\nu_{\alpha}\rightarrow \nu_{\beta}}(t)&=& \frac{1}{3}
    + \frac{1}{2}\left\{ \left[ \rho_{1}^{\alpha}
    \rho_{1}^{\beta}
    \cos\left(\frac{|\Omega_{12}|t}{2}\right)
     + \left(\frac{ \Delta {\cal D}_{21}
   \rho_{1}^{\alpha} \rho_{1}^{\beta} }{|\Omega_{12}|}\right)
    \sin\left(\frac{|\Omega_{12}|t}{2}\right)\right]
    e^{({\cal D}_{11}+{\cal D}_{22})\frac{t}{2}}\right.
    \\ \nonumber &+&\left[ \rho_{4}^{\alpha} \rho_{4}^{\beta}
    \cos\left(\frac{|\Omega_{13}|t}{2}\right) + \left(\frac{\Delta {\cal D}_{54}
    \rho_{4}^{\alpha} \rho_{4}^{\beta} }{|\Omega_{13}|}\right)
    \sin\left(\frac{|\Omega_{13}|t}{2}\right) \right]
    e^{({\cal D}_{44}+{\cal D}_{55})\frac{t}{2}}
    \\ \nonumber &+& \left[ \rho_{6}^{\alpha} \rho_{6}^{\beta}
    \cos\left(\frac{|\Omega_{23}|t}{2}\right)
     +\left(\frac{\Delta {\cal D}_{76}
    \rho_{6}^{\alpha} \rho_{6}^{\beta} }{|\Omega_{23}|}\right)
    \sin\left(\frac{|\Omega_{23}|t}{2}\right) \right]
    e^{({\cal D}_{66}+{\cal D}_{77})\frac{t}{2}}
    \\ \nonumber  &+& \left[ \left(\rho_{3}^{\alpha} \rho_{3}^{\beta}+
     \rho_{8}^{\alpha} \rho_{8}^{\beta}\right)
    \cosh\left(\frac{\Omega_{38}t}{2}\right)\right.
    \\ \nonumber &+& \left.  \left(\frac{2{\cal D}_{38}(\rho_{3}^{\alpha}
     \rho_{8}^{\beta} -
    \rho_{8}^{\alpha} \rho_{3}^{\beta}) + \Delta {\cal D}_{83}
    \left(\rho_{3}^{\alpha} \rho_{3}^{\beta} - \rho_{8}^{\alpha}
    \rho_{8}^{\beta}\right)}{\Omega_{38}}\right)
    \sinh\left(\frac{\Omega_{38}t}{2}\right)\right]
    e^{({\cal D}_{33}+{\cal D}_{88})\frac{t}{2}}~. \\
\label{finalnuprob}
 \ea
Above we have used the notation that $\Delta {\cal D}_{ij}={\cal D}_{ii}-{\cal D}_{jj}$.
We have assumed that
 $2|\Delta_{ij}|<|\Delta {\cal D}_{ij}|$ with the consequence that $\Omega_{ij}$,
$ij=12,13,23$ is imaginary. However,
$\Omega_{38}=\sqrt{({\cal D}_{33}-{\cal D}_{88})^2+4{\cal D}_{38}^{2}}$ will be
real. Taking into account mixing, with the up-to-date values of neutrino mass differences and mixing angles, we can proceed into a fit of (\ref{finalnuprob}) to all the presently available data, including the results of the
neutrino sector of the LSND experiment~\cite{lsnd}, and KamLand spectral distortion data~\cite{kamland}.

\begin{figure}
\centering
\epsfig{figure=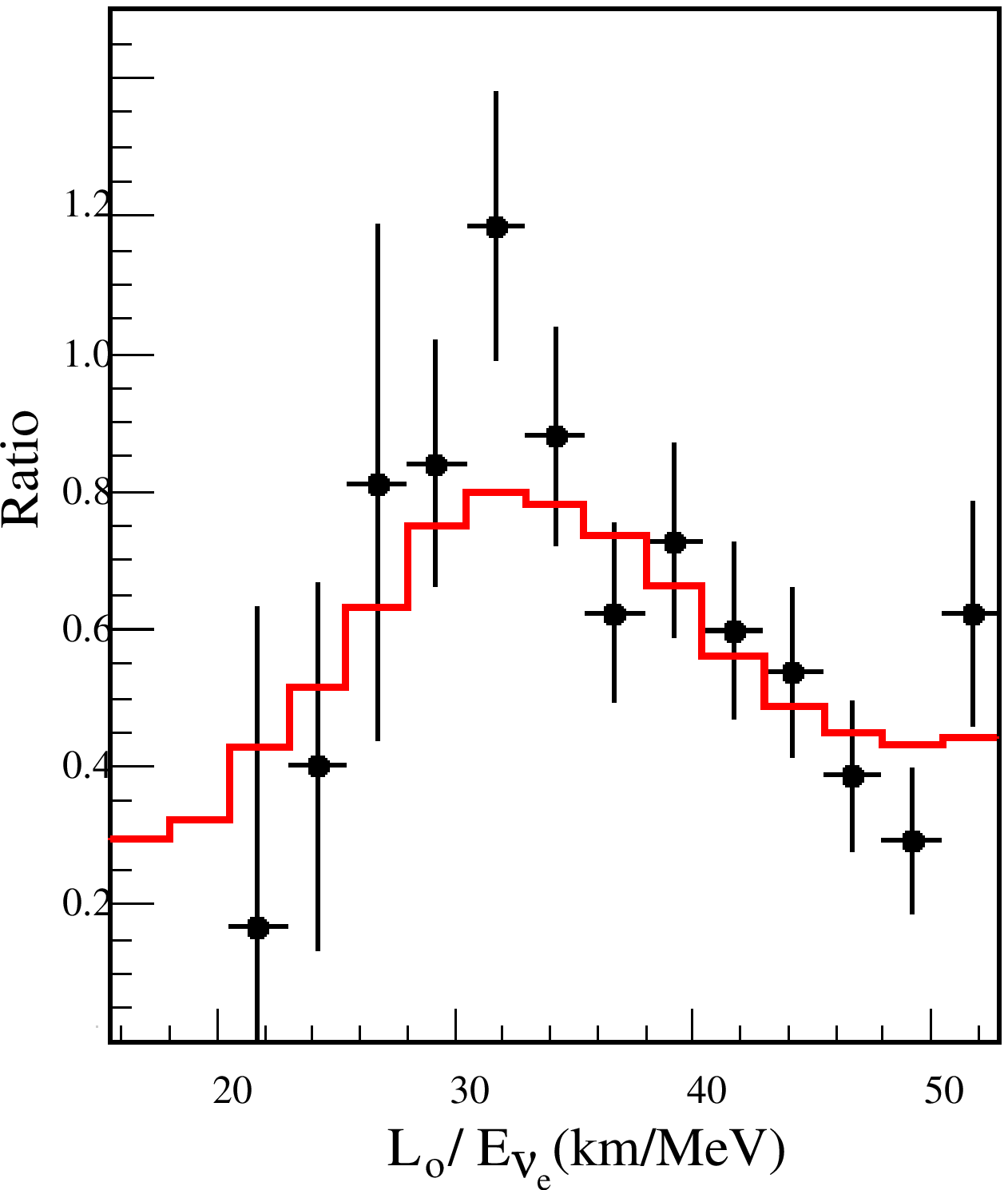,width=5.0cm}\hspace{.5cm}
\epsfig{figure=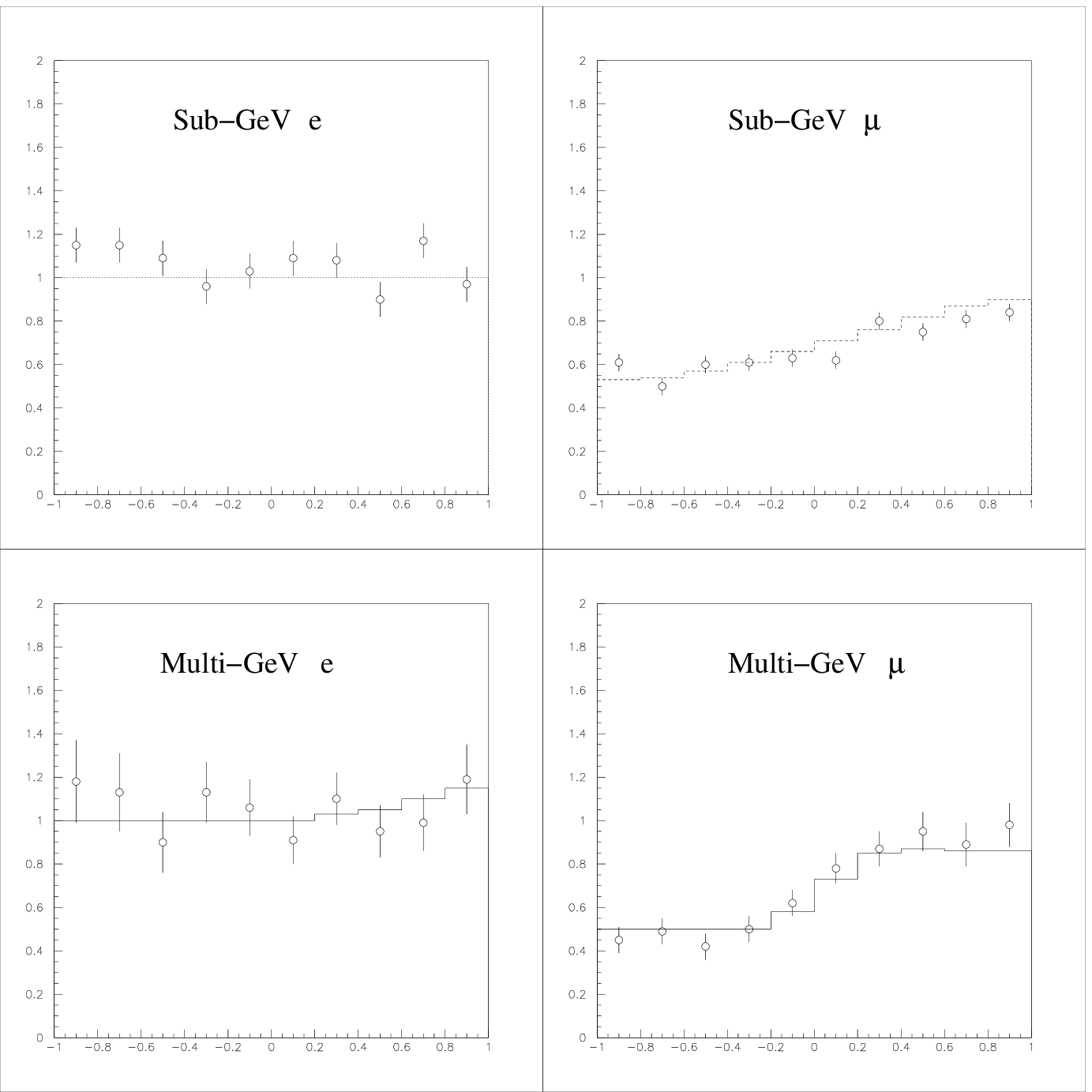,width=6.0cm}
\caption{\underline{Left}: Ratio of the observed $\overline{\nu}_e$ spectrum to the expectation
versus $L_0/E$ for our decoherence model. The dots correspond to KamLAND data.
\underline{Right}:Decoherence fit. The dots correspond to SK data.}
\label{fig1}
\end{figure}

The results are summarised in Fig.~\ref{fig1},
which demonstrates the agreement (left) of our model
with the KamLand spectral
distortion data~\cite{kamland}, and our best fit (right)
for the Lindblad decoherence model used in ref.~\cite{bmsw},
and in
Table \ref{tab:table1}, where we
present the $\chi^2$ comparison for the model in question
and the standard scenario.

\begin{table}[h]
\centering
\begin{tabular}{|c|c|c|}
\hline\hline
$\chi^2$ &  decoherence  & standard scenario  \\ [0.5 ex]
\hline\hline
SK  sub-GeV& 38.0  & 38.2 \\ \hline
SK Multi-GeV & 11.7  & 11.2 \\ \hline
Chooz & 4.5 & 4.5  \\\hline
 KamLAND & 16.7 & 16.6 \\\hline
LSND & 0.  & 6.8  \\\hline
TOTAL & 70.9  & 77.3 \\[1ex]
\hline\hline
\end{tabular}
\caption{$\chi^2$ obtained for our model and the one obtained in the standard
scenario for the different experiments calculated with the same program.}
\label{tab:table1}
\end{table}

The best fit has the feature that only {\it some} of the
oscillation terms in the three generation
probability formula have non trivial damping factors,
with their
exponents being {\it independent} of the oscillation
length,
specifically~\cite{bmsw}. If we denote those non trivial
exponents as ${\cal D}\cdot L$, we
obtain from the best fit of \cite{bmsw}:
\ba
{\cal D}=- \frac{\;\;\; 1.3 \cdot 10^{-2}\;\;\;}
{L},
\label{special}
\ea
in units of 1/km with $L=t$ the oscillation length. The
$1/L$-behaviour of ${\cal D}_{11} $, implies, as we mentioned,
oscillation-length independent Lindblad exponents.

In \cite{bmsw} an analysis of the two types of the theoretical models
of space-time foam, discussed in section \ref{sec:form}, has been performed
in the light of the result of the fit (\ref{special}).
The conclusion was that the model of the
stochastically fluctuating media
(\ref{flct}),(\ref{timedep})
cannot provide the full explanation for the fit, for the following reason:
if the decoherent result of the fit (\ref{special}) was exclusively
due to this model, then the pertinent
decoherent coefficient in that case, for, say, the
KamLand  experiment with an $L \sim 180$~Km,
would be $ |{\cal D}| = \Omega^2 G_N^2 n_0^2 \sim 2.84 \cdot
10^{-21}~{\rm GeV}$ (note that the mixing angle part does not affect the
order of the exponent). Smaller values are found for longer $L$,
such as in atmospheric neutrino experiments or, in future, for high-energy cosmic neutrinos~\cite{icecube}.
The independence of the
relevant damping exponent from the oscillation length, then, as required
by (\ref{special}) may be understood as follows in this context:
In the spirit of \cite{bm2},
the quantity $G_N n_0 = \xi \frac{\Delta m^2}{E}$,
where $\xi \ll 1$ parametrises the contributions of the foam to the
induced neutrino mass differences, according to our discussion
above. Hence, the damping exponent becomes in this case $ \xi^2
\Omega^2 (\Delta m^2)^2 \cdot L /E^2 $. Thus, for oscillation
lengths $L$ we have
$L^{-1} \sim \Delta m^2/E$, and one is left with  the following
estimate for the dimensionless quantity $\xi^2
\Delta m^2 \Omega^2/E \sim 1.3 \cdot 10^{-2}$. This
implies that the quantity $\Omega^2$ is proportional to the
probe energy $E$. In principle,
this is not an unreasonable result, and it is in
the spirit of \cite{bm2}, since back reaction effects onto
space time, which affect the stochastic fluctuations $\Omega^2$, are
expected to increase with the probe energy $E$. However,
due to the smallness of the quantity $\Delta m^2/E$, for energies
of the order of GeV, and $\Delta m^2 \sim 10^{-3}$ eV$^2$,
we conclude (taking into account that
$\xi \ll 1$) that $\Omega^2$ in this case
would be unrealistically large
for a quantum-gravity effect in the model.

We remark at this point that, in such a model,
we can in principle bound independently the $\Omega$
and $n_0$ parameters by looking at the modifications induced by the
medium in the arguments of the oscillatory functions of the
probability (\ref{timedep}), that is the period of oscillation.
Unfortunately this is too small to be detected in the above example,
for which $\Delta a_{e\mu} \ll \Delta_{12}$.

The second model (\ref{flct}),(\ref{timedep}) of stochastic space time
can also be confronted with the data, since in that case
(\ref{special})
would imply for the pertinent damping exponent

\ba
&& \left(\frac{(m_1^2-m^2_2)^2}{2k^2}
   (9\sigma_1+\sigma_2+\sigma_3+\sigma_4)+
\frac{2V\cos2\theta(m_1^2-m_2^2)}{k}
 (12\sigma_1+2\sigma_2-2\sigma_3)
\right)t^2 \nonumber \\
&& \sim 1.3  \cdot 10^{-2}~.
\ea
Ignoring subleading MSW effects $V$, for simplicity,
and considering oscillation lengths $t=L \sim
\frac{2k}{(m_1^2-m^2_2)}$, we then observe that the independence of
the length $L$ result of the experimental fit, found above, may be
interpreted, in this case, as bounding the stochastic fluctuations
of the metric to $9\sigma_1+\sigma_2+\sigma_3+\sigma_4 \sim
1.3. \cdot 10^{-2}$. Again, this is too large to be a quantum gravity
effect, which means that the $L^2$ contributions to the
damping due to stochastic fluctuations of the metric,
as in the model of \cite{ms} above (\ref{flct}),
cannot be the
exclusive explanation of the fit.

The analysis of \cite{bmsw} also demonstrated that, at least as
far as an order of magnitude of the effect is concerned,
a reasonable explanation of the order of the damping
exponent (\ref{special}), is provided by
Gaussian-type energy fluctuations, due to
ordinary physics effects, leading to decoherence-like damping
of oscillation probabilities~\cite{ohlsson}. The order of these fluctuations,
consistent with
the independence of the damping exponent
on $L$ (irrespective of the power of $L$),
is
\ba
 \frac{\Delta E}{E} \sim 1.6 \cdot 10^{-1}
\ea
if one assumes that this is the principal reason for the
result of the fit. However, the {\it selective} damping implied by the result
of the fit  (\ref{special}), implies that this cannot be the explanation.
The fact that the best fit
model includes terms which are not suppressed at all calls for
a more radical explanation of the fit result, and the issue is
still wide open.

It is interesting, however, that the current neutrino data can
already impose stringent constraints on quantum gravity models, and exclude
some of them from being the
exclusive source of decoherence, as we have discussed above.
Of course, this is not a definite conclusion because one cannot
exclude the possibility of other classes of theoretical models
of quantum gravity, which could escape
these constraints. At present, however, we are not aware of any such
theory.

Before closing this section we would like to remark that the above analysis
took into account the neutrino sector results of the LSND experiment~\cite{lsnd}. In the antineutrino sector, the original indication from that experiment was that there was evidence for much more profound oscillations as compared to the neutrino sector, which, if confirmed, should be interpreted as indicating a direct CPT Violation. There were attempts to interpret such a result as being due to either CPT violating mass differences between neutrinos and antineutrinos, characterising, for instance, non-local theories~\cite{lykken},
or CPT-Violating differences between the strengths of the QG-environmental interactions of neutrinos as compared to those of antineutrinos~\cite{barenboim}, without the need for CPT violating mass differences.
In either case, the order of magnitude of the effects indicated by experiment
was too large for the effect to correspond to realistic QG models, probably pointing towards the need of confirming first the LSND results in the antineutrino sector by future experiments before embarking on radical explanations.

\section{Decoherence in Cosmology: some remarks}

\subsection{Cosmic Horizons and Decoherence} 

Recent astrophysical observations, using different experiments
and diverse techniques, seem to indicate that 70\% of the
Universe energy budget is occupied by ``vacuum'' energy density of
unknown origin, termed Dark Energy~\cite{snIa,wmap}.
Best fit models give the positive cosmological {\it constant} Einstein-Friedman
Universe as a good candidate to explain these observations, although
models with a vacuum energy relaxing to zero (quintessential, i.e.
involving a scalar field which has not yet
reached the minimum of its potential) are
compatible with the current data.

{}From a theoretical point of view the two categories of Dark Energy models
are quite different. If there is a relaxing cosmological vacuum energy,
depending on the details of the relaxation rate, it is possible in general
to define asymptotic states and hence a proper scattering matrix (S-matrix)
for the theory, which can thus be quantised canonically.
On the other hand, Universes with a
cosmological {\it constant} $\Lambda > 0$  (de Sitter)
admit no asymptotic states, as a result of the Hubble horizon which
characterises these models, and hampers the definition of proper asymptotic
state vectors, and hence, a proper S-matrix.
Indeed, de Sitter Universes will expand for ever, and eventually
their constant vacuum energy density component will dominate
over matter in such a way that the Universe will enter again an
exponential (inflationary) phase of (eternal) accelerated expansion,
with a Hubble horizon of radius $\delta_H \propto 1/\sqrt{\Lambda}$.
It seems that the recent astrophysical observations~\cite{snIa,wmap}
seem to indicate
that the current era of the universe is the beginning of such an
accelerated expansion.

Canonical quantisation of field theories in de Sitter space times is
still an elusive subject, mostly due to
the above mentioned problem of defining a proper S-matrix.
One suggestion towards the quantisation of such systems  could be
through analogies
with open systems in quantum mechanics, interacting with an environment.
The environment in models with a cosmological constant would consist of field modes
whose wavelength is longer than
the Hubble horizon radius.
This splitting was originally suggested by Starobinski~\cite{staro},
in the context of his stochastic inflationary model, and later on was adopted
by several groups~\cite{coarse}.
Crossing the horizon in either direction
would constitute interactions with the environment.
An initially pure quantum state in such Universes/open-systems would
therefore become eventually mixed, as a result of interactions
with the environmental modes, whose strength will be controlled by the
size of the Hubble horizon, and hence the cosmological constant.
Such decoherent evolution could explain the classicality of the early Universe
phase transitions~\cite{rivers}
(or late in the case of a cosmological constant). The approach is still far from being
complete, not only due to the technical complications,
which force the researchers to adopt severe, and often unphysical
approximations, but also due to conceptual issues, most of which
are associated with the back reaction of matter onto space-time,
an issue often ignored in such a context. It is our opinion
that the latter topic plays an important r\^ole in the evolution
of a quantum Universe, especially one
with a cosmological constant, and is associated
with issues of quantum gravity. The very origin of the cosmological constant,
or in general the dark energy of the vacuum, is certainly a property
of quantum gravity.

This link between quantum decoherence and a cosmological constant
may have far reaching consequences for the phenomenology of
elementary particles, especially neutrinos.
In ~\cite{bm2,ms} we proposed a scenario
according to which the mass differences of neutrinos may have (part of)
their origin in the quantum gravity decoherence medium of space time foam.
The induced decoherence, then, will affect their oscillation,
a notable consequence being the appearance of
intrinsic CPT violating damping terms in front of the oscillation amplitudes.
This fundamental (and local) form of
CPT violation has its origin in the ill-defined nature of
the corresponding  CPT operator
in such decoherent quantum theories,
due to the mathematical theorem by
Wald~\cite{wald}, mentioned previously.
This local form of CPT violation, as a result of the
interaction of the elementary particle with a
decoherent medium
is linked to a cosmological
(global) violation of CPT symmetry of the type proposed in \cite{mlambda}
by means of a generation of a cosmological constant as a result
of neutrino mixing and non-trivial mass differences due to
the quantum gravity vacuum. The framework in which such
a cosmological constant may be generated by the neutrinos is the
approach of \cite{vitiello}, according to which
the problem of mixing
in a quantum field theory is treated by means of a canonical
Fock-space quantization.

\subsection{Non-critical string-framework for Cosmological decoherence}

Let us return to the master equation in (\ref{master})  for non-critical strings. Recent astrophysical observations from different experiments all seem to indicate that $73$\% of the energy of the Universe is in the form of dark
energy. Best fit models give the positive cosmological constant
Einstein-Friedman Universe as a good candidate to explain these
observations. For such de Sitter backgrounds $R_{MN}\propto \Omega g_{MN}$
with $\Omega >0$ a cosmological constant. Also in a perturbative derivative
expansion (in powers of $\alpha ^{\prime }$ where $\alpha ^{\prime
}=l_{s}^{2}$ is the Regge slope of the string and $l_{s}$ is the fundamental
string length) in leading order
\begin{equation}
\beta _{\mu \nu }=\alpha ^{\prime }R_{\mu \nu }=\alpha ^{\prime }\Omega
g_{\mu \nu }  \label{fixedpoint}
\end{equation}%
and
\begin{equation}
{\cal G}_{ij}=\delta _{ij}.
\end{equation}%
This leads to
\begin{equation}
\partial _{t}\rho =i\left[ \rho ,H\right] +\,\alpha ^{\prime }\Omega :g_{MN}%
\left[ g^{MN},\rho \right] :  \label{master2}
\end{equation}%
For a weak-graviton expansion about flat space-time, $g_{MN}=\eta
_{MN}+h_{MN}$, and
\begin{equation}
h_{0i}\propto \frac{\Delta p_{i}}{M_{P}}.  \label{recoil2}
\end{equation}%
If an antisymmetric ordering prescription is used, then the master equation
for low energy string matter assumes the form%
\begin{equation}
\stackrel{.}{\partial _{t}\rho _{Matter}}=i\left[ \rho _{Matter},H\right]
-\,\Omega \left[ h_{0j},\left[ h^{0j},\rho _{Matter}\right] \right]
\label{master3}
\end{equation}%
( when $\alpha ^{\prime }$ is absorbed into $\Omega )$. In view of the
previous discussion this can be rewritten as%
\begin{equation}
\stackrel{.}{\partial _{t}\rho _{Matter}}=i\left[ \rho _{Matter},H\right]
-\,\Omega \left[ \overline{u}_{j},\left[ \overline{u}^{j},\rho _{Matter}%
\right] \right] ~.  \label{master4}
\end{equation}%
thereby giving the {\it master equation for Liouville decoherence} in the
model of a D-particle foam with a cosmological constant.

The above D-particle inspired approach deals with possible non-perturbative
quantum effects of gravitational degrees of freedom. The analysis is
distinct from the phenomenology of dynamical semigroups which does not
embody specific properties of gravity. Indeed the phenomenology is
sufficiently generic that other mechanisms of decoherence such as the MSW
effect~\cite{msw} can be incorporated within the same framework. Consequently an
analysis which is less generic and is related to the specific \ decoherence
implied by non-critical strings is necessary.It is sufficient to study a
massive non-relativistic particle propagating in one dimension to establish
qualitative features of D-particle decoherence. The environment will be
taken to consist of both gravitational and non-gravitational degrees of
freedom; hence we will consider a generalisation of quantum Brownian motion
for a particle which has additional interactions with D-particles. This will
allow us to compare qualitatively the decoherence due to different
environments.The non-gravitational degrees of freedom in the environment (in
a thermal state) are conventionally modelled by a collection of harmonic
oscillators with masses $m_{n}$, frequency $\omega _{n}$ and co-ordinate
operator $\widehat{q}_{n}$ coupled to the particle co-ordinate $\widehat{x}$
by an interaction of the form $\sum_{n}g_{n}\widehat{x}\widehat{q}_{n}$. The
master equation which is derived can have time dependent coefficients due to
the competing timescales, e.g. relaxation rate due to coupling to the
thermal bath, the ratio of the time scale of the harmonic oscillator to the
thermal time scale etc. However an ab initio calculation of the
time-dependence is difficult to do in a rigorous manner. It is customary to
characterise the non-gravitational environment by means of its spectral
density $I\left( \omega \right) \left( =\sum_{n}\delta \left( \omega -\omega
_{n}\right) \frac{g_{n}^{2}}{2m_{n}\omega _{n}}\right) $. The existence of
the different time scales leads in general to non-trivial time dependences
in the coefficients in the master equation which are difficult to calculate
in a rigorous manner \cite{BLH1992}. The dissipative term in (\ref{master4})
involves the momentum transfer operator due to recoil of the particle from
collisions with D-particles (\ref{recoil}). This transfer process will be
modelled by a classical Gaussian random variable $r$ which multiplies the
momentum operator $\widehat{p}$ for the particle:
\begin{equation}
\overline{u_{i}}\qquad \rightarrow \qquad \frac{r}{M_{P}}\widehat{p}
\label{trnsf}
\end{equation}%
Moreover the mean and variance of $r$ are given by
\begin{equation}
\left\langle r\right\rangle =0~,\qquad {\rm and}\qquad \left\langle
r^{2}\right\rangle =\sigma ^{2}~.  \label{random}
\end{equation}%
On amalgamating the effects of the thermal and D-particle environments, we
have for the reduced master equation \cite{msdark} for the matter (particle)
density matrix $\rho $ (on dropping the Matter index)%
\begin{equation}
i\frac{\partial }{\partial t}\rho =\frac{1}{2m}\left[ \widehat{p}^{2},\rho %
\right] -i\Lambda \left[ \widehat{x},\left[ \widehat{x},\rho \right] \right]
+\frac{\gamma }{2}\left[ \widehat{x},\left\{ \widehat{p},\rho \right\} %
\right] -i\Omega r^{2}\left[ \widehat{p},\left[ \widehat{p},\rho \right] %
\right]  \label{master5}
\end{equation}%
where $\Lambda ,\gamma $ and $\Omega $ are real time-dependent coefficients.
As discussed in \cite{msdark} a possible model for $\Omega \left( t\right) $
is
\begin{equation}
\Omega \left( t\right) =\Omega _{0}+\frac{\widetilde{\gamma }}{a+t}+\frac{%
\widetilde{\Gamma }}{1+bt^{2}}  \label{cosmological}
\end{equation}%
where $\omega _{0}$, $\widetilde{\gamma }$, $a$, $\widetilde{\Gamma }$ and $%
b $ are positive constants. The quantity $\widetilde{\gamma }<1$ contains
information on the density of D-particle defects on a four-dimensional
world.The time dependence of $\gamma $ and $\Lambda $ can be calculated in
the weak coupling limit for general $n$ (i.e. ohmic, $n=1$ and non-ohmic $%
n\neq 1$ environments)$\ $where
\begin{equation}
I\left( \omega \right) =\frac{2}{\pi }m\gamma _{0}\omega \left[ \frac{\omega
}{\varpi }\right] ^{n-1}e^{-\omega ^{2}/\varpi ^{2}}  \label{spectral}
\end{equation}%
and $\varpi $ is a cut-off frequency. The precise time dependence is
governed by $\Lambda \left( t\right) =\int_{0}^{t}ds\,\nu \left( s\right) $
and $\gamma \left( t\right) =\int_{0}^{t}ds\,\nu \left( s\right) s$ where $%
\nu \left( s\right) =\int_{0}^{\infty }d\omega \,I\left( \omega \right)
\coth \left( \beta \hbar \omega /2\right) \cos \left( \omega s\right) $. For
the ohmic case, in the limit $\hbar \varpi \ll k_{B}T$ followed by $\varpi
\rightarrow \infty $, $\Lambda $ and $\gamma $ are given by $m\gamma
_{0}k_{B}T$ and $\gamma _{0}$ respectively after a rapid initial transient.
For high temperatures $\Lambda $ and $\gamma $ have a powerlaw increase with
$t$ for the subohmic case whereas there is a rapid decrease in the
supraohmic case.

This formalism can be applied to study flavour oscillations in the 
presence of such dark-energy induced decoherence~\cite{msdark}. 
With the general $\Omega \left( t\right) $
of (\ref{cosmological}) the probability for flavour oscillation, $%
P_{1\rightarrow 2}$, is found proportional to
\begin{equation}
P_{1\rightarrow 2}\propto \left( \sin 2\theta \right) ^{2}\left( 1-\exp
\left[ -4\sigma ^{2}p^{2}\mathcal{I}(t)\right] \right) \times [{\rm conventional~oscillatory~terms}]~.  \label{flavourprob}
\end{equation}%
with $\mathcal{I}(t)\equiv \int_{0}^{t}\Omega (t^{\prime })dt^{\prime }=\Omega
_{0}t+\widetilde{\gamma }\mathrm{ln}(1+t/a)+\frac{\widetilde{\Gamma }}{\sqrt{%
b}}\mathrm{tan}^{-1}(\sqrt{b}t)$~.

{}From (\ref{flavourprob}) it is evident that decoherence affects
this probability with an exponential damping \emph{only if} the cosmological
term $\Omega $ is the \emph{constant }$\Omega _{0}$. In particular, in the
absence of the $\Omega _{0}$ term, and in the limit $\widetilde{\Gamma }=0$,
the decoherence due to the D-particle foam results in power damping for
large times $t\rightarrow \infty $, the terms $\widetilde{\rho }_{i}\sim
t^{-\delta _{i}^{2}}$, $i=0,3$, while $\widetilde{\rho }_{i}\sim
t^{-p^{2}\delta _{i}^{2}}$, $i=1,2$, i.e. the scaling power depends on the
probe's momentum (with $\delta _{i}$ appropriate constants depending on the
term we look at). 

Although tiny, for laboratory scales, such cosmological decoherence 
effects may play an important r\^ole in the propagation of 
sensitive particle physics probes, such as high energy neutrinos, over 
cosmological distances, especially in theories where the gravity scale may be at TeV~\cite{anchor}. 

\subsection{Non-Critical String effects on relic abundances and dark matter constraints}

Before closing this section we would like to make a final remark on contributions to the Boltzmann equation (determining cosmic relic abundances) from non-critical
string terms, proportional to the $\beta$-functions $\beta^i$. 
Such quantities are important
for constraining dark matter (in particular supersymmetric) candidates, ,  
from astrophysical data~\cite{wmap,susyconstr}.

In effective four-dimensional space-times (after appropriate string compactification of the extra dimensions) the relic density
of a species $\lsp$ of mass $m_{\lsp}$, assumed to be the 
dominant dark matter candidate, is given generically by~\cite{lmn}:
\ba
\Omega_{\tilde{\chi}} h_0^2 \;=\; \left( \Omega_{\tilde{\chi}} h_0^2 \right)_{no-source}
\times {\left(  \frac{{\tilde g}_{*}}{g_{*}}   \right)}^{1/2} \;
{\rm exp}\left( \int_{x_{0}}^{x_f}  \frac{\Gamma H^{-1}}{x} dx \right) \quad . \label{relic}
\ea
In the above formula, 
$H$ is the Hubble parameter of the Universe, $x \equiv T/m_{\lsp}$, with $T$ the temperature, and
$\left( \Omega_{\tilde{\chi}} h_0^2 \right)_{no-source}\;=\;\
\frac{1.066\; \times  10^9\; {\rm GeV}^{-1}}{ M_{Planck}
\;\sqrt{g_{*}} \;J} $, ~$J\equiv \int_{x_0}^{x_f} \vev{ v \sigma} dx$,
is the standard-cosmology result without sources~\cite{kolb}. 
The source $\Gamma $ contains also, in our approach, {\it both} time-dependent dilaton and non-critical-string terms:
$\Gamma (t) \equiv  \dot \Phi  - \frac{1}{2}e^{-\Phi}g^{\mu\nu}{\tilde \beta}_{\mu\nu}^{\rm Grav}~,$,
where the suffix ``Grav'' denotes graviton background $\beta$-functions. The time-dependent dilaton sources $\Phi (t)$ could play the r\^ole of quintessence fields in string cosmology~\cite{veneziano} and be responsible for the Universe's current acceleration (through dilatonic dark energy which relaxes asymptotically to zero).

The freeze out point $x_f$ (defined to be the temperature below which all (local) particle interactions are frozen due to the expansion of the Universe)
contains source-term modifications:
\ba
x_f^{-1}\;=\; ln \left[ 0.03824 \; g_s\; \frac{M_{Planck}~
m_{\lsp}}{\sqrt{g_{*}}} x_f^{1/2} {\vev{ v \sigma}}_f \right]\;+\;
\frac{1}{2} \;  ln \left( \frac{g_{*}}{ {\tilde{g}}_{*} }\right)
\;+\; \int_{x_f}^{x_{in}} \;  \frac{\Gamma H^{-1}}{x} \;dx ~.
\label{frpo}
\ea
whereby~\cite{lmn} ${\tilde g}_{eff}=g_{eff} + \frac{30}{\pi^2} T^{-4} \Delta \rho $, with $g_{eff}$ the ordinary matter degrees of freedom, assumed thermal.

The merit of casting the relic density in such a form is that it clearly
exhibits the effect of the presence of the source.
It is immediately seen from (\ref{relic}) that, depending on the sign of the source $\Gamma$, one may have increase or reduction of the relic density as compared with the corresponding value in the absence of the source. 
For instance, for simple de-Sitter backgrounds, with positive cosmological constant $\Omega_0 > 0$, and constant
dilaton, $\Phi = \phi_0$, the source term $\Gamma$ in (\ref{relic}) is 
simply 
$\Gamma \sim -2e^{\-\phi_0}\Omega_0 $, in (effective) four space-time dimensions, to lowest order in the Regge slope $\alpha '$ (c.f. \ref{fixedpoint}), and $H^{-1} \Gamma  \propto -e^{-\phi_0}$. However,
in realistic string cosmologies with matter, there may be time-dependent dilatons and relaxation dark-energy components present~\cite{veneziano}, and thus the situation is more complicated~\cite{lmn}.
It becomes evident from the above that the constraints imposed on 
physically appealing particle physics models, such as 
supersymmetric extensions of the Standard Model, from astrophysical
bounds~\cite{wmap} on, say, cold dark matter relic densities~\cite{susyconstr}, need to be revisited in the non-critical (decoherent) string context~\cite{lmn}.

\section{Conclusions and Outlook}

In this review we have outlined several aspects of decoherence-induced CPTV and
the corresponding experiments.
We described the interesting and challenging precision tests
that can be performed using kaon systems, especially $\phi$-meson factories,
where some unique ($\omega$) effects on EPR-correlation modifications, associated with the ill-defined nature of CPT
operator in decoherent QG, may be in place.

We have presented sufficient theoretical motivation and
estimates of the associated effects to support the case that
testing QG experimentally at present
facilities may turn out to be a worthwhile endeavour.
In fact, as we have argued, CPTV may be a real feature of QG,
that can be tested and observed, if true, in the foreseeable future.
Indeed, as we have seen, some theoretical (string-inspired) models of
space-time foam predict $\omega$-like effects of an order of
magnitude that is already well within the reach of the next upgrade of
$\phi$-factories, such as DA$\Phi$NE-2.
Neutrino systems is another extremely sensitive probe of particle physics
models which can already falsify several models or place stringent
bounds in others.
Finally, QG-decoherence effects in Cosmology may also play a r\^ole in our understanding of the Universe evolution, or even modify the current astrophysical constraints on models of particle physics, where such possible decoherent effects are ignored.

Therefore, the current experimental situation for QG
signals appears exciting,
and several experiments are reaching interesting regimes, where many
theoretical models can be falsified. More precision experiments are
becoming available, and many others are being designed for the
immediate future. Searching for tiny effects of this elusive theory
may at the end be very rewarding. Surprises may be around the
corner, so it is worth investing time and effort.

\section*{Acknowledgements}

The authors would like to thank the organisers of DICE2006 Conference (Piombino, Italy, 11-15 September 2006)
for the invitation. This work was supported in part by the European Union through the Marie Curie Research and Training Network UniverseNet (MRTN-CT-2006-035863).

\paragraph{}

\end{document}